\documentclass[longauth]{aa}
\usepackage{graphicx}
\usepackage{txfonts}
\usepackage{lscape,longtable}
\usepackage{natbib}
\bibpunct{(}{)}{,}{a}{}{,}

\begin{document}

\title{Gaia-ESO Survey: Role of magnetic activity and starspots on
pre-main-sequence lithium evolution%
\thanks{Tables B.1-B.6 are only available in electronic form at the CDS via
anonymous ftp to cdsarc.u-strasbg.fr (130.79.128.5) or via
http://cdsweb.u-strasbg.fr/cgi-bin/qcat?J/A+A/}%
\fnmsep\thanks{Based on observations collected with ESO telescopes at the La
Silla Paranal Observatory in Chile, for the Gaia-ESO Large Public
Spectroscopic Survey (188.B-3002, 193.B-0936, 197.B-1074)}
}

\author{E. Franciosini\inst{\ref{oaa}}
\and E. Tognelli\inst{\ref{fzu}}
\and S. Degl'Innocenti\inst{\ref{unipi},\ref{infnpi}}
\and P.~G. Prada~Moroni\inst{\ref{unipi},\ref{infnpi}}
\and S. Randich\inst{\ref{oaa}}
\and G.~G. Sacco\inst{\ref{oaa}} 
\and L. Magrini\inst{\ref{oaa}} 
\and E. Pancino\inst{\ref{oaa},\ref{ssdc}} 
\and A.~C. Lanzafame\inst{\ref{unict}}  
\and R. Smiljanic\inst{\ref{copernicus}} 
\and L. Prisinzano\inst{\ref{oapa}} 
\and N. Sanna\inst{\ref{oaa}}
\and V. Roccatagliata\inst{\ref{unipi},\ref{infnpi},\ref{oaa}}
\and R. Bonito\inst{\ref{oapa}}
\and P. de~Laverny\inst{\ref{nice}}
\and M.~L. Guti\'errez Albarr\'an\inst{\ref{ucm}}
\and D. Montes\inst{\ref{ucm}}
\and F. Jim\'enez-Esteban\inst{\ref{csic}}
\and G. Gilmore\inst{\ref{cambridge}} 
\and M. Bergemann\inst{\ref{mpifa}}
\and G. Carraro\inst{\ref{unipd}} 
\and F. Damiani\inst{\ref{oapa}}
\and A. Gonneau\inst{\ref{cambridge}}
\and A. Hourihane\inst{\ref{cambridge}} 
\and L. Morbidelli\inst{\ref{oaa}} 
\and C.~C. Worley\inst{\ref{cambridge}} 
\and S. Zaggia\inst{\ref{oapd}}
}

\institute{%
INAF - Osservatorio Astrofisico di Arcetri, Largo E. Fermi 5, 50125,
Florence, Italy \label{oaa} 
\email{elena.franciosini@inaf.it}
\and
CEICO, Institute of Physics of the Czech Academy of Sciences, Na Slovance
2, 182 21 Praha 8, Czechia\label{fzu}
\and
Department of Physics "E. Fermi", University of Pisa, Largo Bruno
Pontecorvo 3, 56127, Pisa, Italy \label{unipi}
\and
INFN, Section of Pisa, Largo Bruno Pontecorvo 3, 56127, Pisa, Italy
\label{infnpi}
\and
Space Science Data Center - Agenzia Spaziale Italiana, Via del Politecnico
s.n.c., I-00133 Roma, Italy \label{ssdc}
\and
Dipartimento di Fisica e Astronomia, Sezione Astrofisica, Universit\`a di
Catania, via S. Sofia 78, 95123, Catania, Italy \label{unict}
\and
Nicolaus Copernicus Astronomical Center, Polish Academy of Sciences, ul.
Bartycka 18, 00-716, Warsaw, Poland \label{copernicus}
\and
INAF - Osservatorio Astronomico di Palermo, Piazza del Parlamento 1, 90134,
Palermo, Italy \label{oapa}
\and
Laboratoire Lagrange (UMR7293), Universit\'e de Nice Sophia Antipolis,
CNRS, Observatoire de la C\^ote d'Azur, CS 34229, F-06304 Nice cedex 4,
France
\label{nice}
\and
Departamento de F\'{\i}sica de la Tierra y Astrof\'{\i}sica and
IPARCOS-UCM, Instituto de F\'{\i}sica de Part\'{\i}culas y del Cosmos de
la UCM, Facultad de Ciencias Físicas, Universidad Complutense de Madrid,
28040 Madrid, Spain
\label{ucm}
\and 
Departamento de Astrof\'{\i}sica, Centro de Astrobiolog\'{\i}a
(CSIC-INTA), ESAC Campus, Camino Bajo del Castillo s/n, E-28692 Villanueva
de la Ca\~nada, Madrid, Spain \label{csic}
\and
Institute of Astronomy, University of Cambridge, Madingley Road, Cambridge
CB3 0HA, United Kingdom \label{cambridge}
\and
Max-Planck Institut f\"{u}r Astronomie, K\"{o}nigstuhl 17, 69117
Heidelberg, Germany \label{mpifa}
\and
Dipartimento di Fisica e Astronomia, Universit\`a di Padova, Vicolo
dell'Osservatorio 3, 35122 Padova, Italy \label{unipd}
\and
INAF - Osservatorio Astronomico di Padova, Vicolo dell'Osservatorio 5,
35122 Padova, Italy \label{oapd}
}

\date{Received ... / Accepted ...}
\titlerunning{Role of starspots on the pre-main-sequence lithium evolution}

\abstract{
It is now well-known that pre-main-sequence models with inflated radii
should be taken into account to simultaneously reproduce the
colour-magnitude diagram and the lithium depletion pattern observed in young
open star clusters.
}{
We tested a new set of pre-main-sequence models that include radius
inflation due to the presence of starspots or to magnetic inhibition of
convection. We used five clusters observed by the Gaia-ESO Survey that span
the age range $\sim$\,10\,--\,100~Myr, in which these effects could be
important.
}{
The Gaia-ESO Survey radial velocities were combined with astrometry from
{\em Gaia}~EDR3 to obtain clean lists of high-probability members for the
five clusters. A Bayesian maximum likelihood method was adopted to fit the
observed cluster sequences to theoretical predictions to derive the best
model parameters and the cluster reddening and age. Models were calculated
with different values of the mixing length parameter
($\alpha_\mathrm{ML}=2.0$, 1.5 and 1.0) for the cases without spots or with
effective spot coverage $\beta_\mathrm{spot}=0.2$ and 0.4. The models were
also compared with the observed lithium depletion patterns.
}{
To reproduce the colour-magnitude diagram and the observed lithium depletion
pattern in Gamma~Vel~A and B and in 25~Ori, both a reduced convection
efficiency, with $\alpha_\mathrm{ML}=1.0$, and an effective surface spot
coverage of about 20\% are required. We obtained ages of
18$^{+1.5}_{-4.0}$~Myr and 21$^{+3.5}_{-3.0}$~Myr for Gamma~Vel~A and B,
respectively, and $19^{+1.5}_{-7.0}$~Myr for 25~Ori. However, a single
isochrone is not sufficient to account for the lithium dispersion, and an
increasing level of spot coverage as mass decreases seems to be required. On
the other hand, the older clusters (NGC~2451\,B at 30$^{+3.0}_{-5.0}$~Myr,
NGC~2547 at 35$^{+4.0}_{-4.0}$~Myr, and NGC~2516 at 138$^{+48}_{-42}$~Myr)
are consistent with standard models (i.e. $\alpha_\mathrm{ML}=2.0$ and no
spots) except at low masses: a 20\% spot coverage appears to reproduce the
sequence of M-type stars better and might explain the observed spread in
lithium abundances.
}{
The quality of Gaia-ESO data combined with {\em Gaia} allows us to gain
important insights on pre-main-sequence evolution. Models including
starspots can provide a consistent explanation of the cluster sequences and
lithium abundances observed in young clusters, although a range of starspot
coverage is required to fully reproduce the data.
}


\maketitle

\section{Introduction}
\label{sec:intro}

Lithium is a key tracer of mixing processes in stellar interiors: since it
is burned at relatively low temperatures ($\sim\,$2.5-3~MK), its surface
abundance is rapidly depleted in low-mass pre-main-sequence (PMS) stars
with deep convective envelopes at a rate that is dependent on mass 
\citep[][and references therein]{rm21,tognelli21b}. Fully convective stars
start to deplete lithium at an age of 5--10~Myr, and they destroy it
completely in a few dozen million years \citep{bildsten97}. However, lithium
is fully preserved in very low-mass stars and brown dwarfs that do not reach
the lithium-burning temperatures at their centres, creating a sharp
transition between lithium-depleted and undepleted stars that is called the
`lithium depletion boundary' \citep[LDB;
e.g.][]{dantona94,basri96,stauffer00,burke04,tognelli15b}. The position of
the LDB is strongly dependent on age and moves towards lower masses as age
increases, but it only weakly depends on other parameters. The LDB therefore
constitutes a robust age indicator for young stellar associations and open
clusters from $\sim$\,20--30~Myr \citep[e.g.][]{manzi08,jeffries13} up to
the age of the Hyades \citep{martin18}. Moreover, for higher-mass stars that
develop a radiative core during the PMS, the depletion of the surface
lithium abundance in this phase depends on the temperature profile and on
the sink of the external convection, and thus also on the mass and chemical
composition \citep[e.g.][and references
therein]{ventura98,tognelli12,baraffe15,baraffe17}. For these reasons, the
analysis of surface lithium abundances in young open clusters of different
ages can provide important constraints on stellar evolutionary models. 

The Gaia-ESO Survey \citep[hereafter GES;][]{GES12,GES13} is particularly
well suited for these investigations since it provides the largest database
of homogeneously determined stellar parameters and lithium abundances for
several open clusters, spanning a large age range from a few million to
several billion years. GES observations also place constraints on the
post-main-sequence evolution and on the origin of the Li-rich giant stars
\citep{casey16,smiljanic18,magrini21}.

\citet{jeffries17} compared the GES observations of the young open cluster
Gamma~Vel with the \citet{baraffe15} and Darthmouth \citep{dotter08}
standard evolutionary models. They found that these models are unable to
simultaneously reproduce the colour-magnitude diagram (CMD) and the
lithium depletion pattern: while the CMD could be well fitted by standard
models with an age of $\sim$7.5~Myr, the strong lithium depletion observed
in M-type stars implies a significantly older age, and its pattern is
inconsistent with standard model predictions. A similar discrepancy
between standard models and the observed lithium depletion pattern was
also noted by \citet{messina16} in the $\beta$~Pictoris moving group.
However, \citet{jeffries17} were able to reconcile the lithium pattern and
the CMD at a common age of $\sim 18-21$~Myr by assuming that the radius of
low-mass stars is inflated by $\sim 10$\%, and by accordingly modifying
the models considering (almost) fully convective stars with a simple
polytropic structure. At fixed mass and age, a radius inflation results in
a significantly reduced effective temperature, leading to a reduced
central temperature and thus to a delayed lithium depletion during the PMS
phase. This results in significant older ages than in the standard
scenario when the CMD is compared to model isochrones, and in a
significant shift of the centre of the lithium depleted region toward
lower effective temperatures. A non-uniform radius inflation in the PMS
phase could also explain the observed dispersion of the surface lithium
abundance in some young clusters \citep[see e.g.][]{somers15}.

Evidence of inflated radii and reduced effective temperatures with respect
to model predictions has been found in close, tidally locked eclipsing
binaries \citep[see e.g.][and references therein]{torres13}, and in single
active low-mass stars both in the field
\citep[e.g.][]{morales08,kesseli18} and in open clusters
\citep[e.g.][]{ventura98,jackson09,tognelli12,jackson16,jackson18,somers17},
suggesting that radius inflation may be linked to magnetic activity.
Further support to this idea was provided by \citet{jaehnig19}, who found
that all inflated stars in the Hyades and Pleiades have a Rossby number
$R_\circ \la 0.1$, corresponding to saturation of magnetic activity
\citep[e.g.][]{pizzol03}. \citet{jeffries21} showed that radius inflation
from magnetic models is able to reproduce the upper envelope of the
lithium pattern observed in the 120~Myr old M35 cluster, and confirmed the
correlation between faster rotation, radius inflation, and reduced lithium
depletion that was noted by \citet{somers17} in the Pleiades.
\citet{binks21} found that models with high levels of magnetic activity
are required to reconcile the LDB age of NGC~2232 with the age derived
from isochrone fitting.

Strong magnetic fields can affect the stellar structure by reducing the
convection efficiency in the external superadiabatic region, which can be
simulated by a reduction of the mixing length parameter \citep[][and
references therein]{chabrier07,feiden12,feiden13,feiden14}. However, the
size of this effect depends on stellar mass: low-mass fully convective
stars, which are almost fully adiabatic, are less sensitive to variations in
the mixing length than more massive stars, implying that unreasonably high
magnetic fields would be needed to significantly modify the characteristics
of fully convective stars. Another phenomenon linked to strong magnetic
fields is a large coverage of the stellar surface by cool starspots that
block the emerging energy flux, reducing the stellar radiation, and
consequently causing the star to inflate
\citep[e.g.][]{chabrier07,macdonald13,jackson14,somers15,somers20}.

In this paper, we further explore this issue in a consistent way, using a
set of stellar models that was specifically computed for our analysis.  In
particular, we explore the effects of the presence of starspots adopting a
fully consistent treatment, that is, we include these effects in the Pisa
stellar evolutionary code \citep[see e.g.][]{tognelli11} and follow the
resulting evolution of lithium depletion with age. We also adopt for our
comparison PMS models with lower external convection efficiency (i.e.
lower mixing length values) to search for a better agreement mainly for
higher-mass, not fully convective stars, which are more sensitive to
variations in the convection efficiency.

We apply our analysis to five open clusters of ages between $\sim\,$10 and
100~Myr that were observed within GES (\object{25~Ori}, \object{Gamma~Vel},
\object{NGC~2547}, \object{NGC~2451\,B,} and \object{NGC~2516}). These
clusters were selected because they cover the age interval in which the
effect of radius inflation could be significant, allowing us to investigate
how it evolves with age. 
The 25~Ori cluster is a group of PMS stars that was discovered by
\citet{briceno05} in the Orion OB1a association, with an estimated age of
6-13~Myr \citep{downes14,briceno19,kos19,zari19}; a dispersed, kinematically
distinct population was also found in the region using data from the {\em
Gaia} Second Data Release \citep[DR2; e.g.][]{zari19}. Because only a few
stars of the secondary population were observed by GES, only the main
cluster is considered here.
Gamma~Vel \citep[age $\sim$\,10--20~Myr,][]{jeffries14,jeffries17} and
NGC~2547 \citep[$35 \pm 3$~Myr,][]{jeffries05} are both located in the
Vela~OB2 association at a relative separation of $\sim\,$2~degrees. Both
clusters host two kinematically distinct populations
\citep{jeffries14,sacco15}; the two Gamma~Vel populations (Gamma~Vel~A and
B) are also separated by $\sim\,$38~pc along the line of sight
\citep{francio18}.
NGC~2451 is a double cluster composed of two open clusters of similar age
\citep[30$-$40~Myr,][]{randich18} located at different distances along the
same line of sight \citep{roeser94,platais96}. The GES observations cover
the background cluster NGC~2451\,B and only a few selected regions of the
closer and more dispersed NGC~2451\,A. For this reason, we considered only
NGC~2451\,B here.
Finally, NGC~2516 is the oldest cluster in our sample, with an age of
$\sim\,100-140$~Myr \citep[e.g.][]{lyra06,randich18}, so that most of its
members are already close to or at their main-sequence position. 
All clusters have solar or slightly subsolar metallicities
\citep{biazzo11,jacobson16,spina17}.

Three of our sample clusters (NGC~2451\,B, NGC~2547, and NGC~2516) were
included in the study by \citet{randich18}, who combined the GES results
with the parallaxes and proper motions from the Tycho-Gaia Astrometric
Solution (TGAS) to derive new ages and reddening values from isochrone
fitting, using different recent sets of standard theoretical models,
including those adopted here. 
In this paper we improve the membership selection using the parallaxes and
proper motions from the {\em Gaia} Early Data Release 3 (EDR3), and we
verify whether the ages from \citet{randich18} agree with those derived from
the surface lithium abundances. We also verify whether the addition of
magnetic effects can improve the results.

The paper is organised as follows: in Sect.~\ref{sec:data} we describe the
GES observations and {\em Gaia} data, and the membership selection is
described in Sect.~\ref{sec:membership}. The theoretical models and their
comparison with the observations are described in
Sect.~\ref{sec:model_comp}. Discussion and conclusions are given in
Sect.~\ref{sec:disc}.

\section{Data}
\label{sec:data}

The GES observations were carried out with the FLAMES multi-object
spectrograph mounted on the VLT/UT2 telescope \citep{pasquini02}, using the
Giraffe instrument operated in MEDUSA mode and the high-resolution UVES
spectrograph ($R=47\,000$). In the case of young clusters, the Giraffe HR15N
grating (644--680~nm, $R\sim\,17\,000$) and the UVES U580 (480--680~nm) and
U520 (420--620~nm) setups were used. The HR15N and U580 setups include the
\ion{Li}{i} doublet at 6707.8~\AA{} as well as the H$\alpha$ line. The
selection of targets was based on photometry only, and was performed in an
homogeneous way for all clusters, keeping all objects that fall inside a
sufficiently wide band around the cluster locus in the available optical and
near-infrared CMDs to ensure that all possible cluster members are included
without introducing any biases \citep[see][for more details]{bragaglia21}.
For young clusters, UVES fibres were allocated to the brightest stars,
choosing preferentially previously known cluster members if the
information was available. Some of the targets were observed with both
UVES and Giraffe for cross-calibration purposes.

We used the recommended radial velocities (RVs), atmospheric parameters, and
lithium abundances contained in the Sixth Internal Data Release (iDR6). A
detailed description of the data reduction pipelines and the derivation of
radial and rotational velocities is provided by \citet{gilmore21}
and \citet{sacco14} for Giraffe and UVES, respectively. A summary of the
Giraffe data reduction is also presented in \citet{jeffries14}.
Atmospheric parameters were derived by several analysis teams within
different working groups, depending on stellar-type and/or age and/or
instrumental setup. The results were then homogenised following the
calibration strategy described by \citet{pancino17} and combined to produce
the final recommended values and the corresponding statistical uncertainties
\citep[see][Worley et al., in prep., Hourihane at al. in prep. for
details]{smiljanic14,lanzafame15,blomme21}. 
Lithium abundances $A$(Li)\footnote{%
In the usual notation $A$(Li)$=\log N(\mathrm{Li})/N(\mathrm{H})+12$.} 
for FGK stars in iDR6 were derived by measuring the equivalent width (EW) of
the lithium doublet at 6707.8~\AA{} using a Gaussian fit of each component.
Abundances were then computed using a new set of curves of growth that was
specifically derived for GES (Franciosini et al., in prep.). The curves of
growth were measured on a grid of synthetic spectra that covers the whole
parameter space of the survey and that was computed as in
\citet{delaverny12} and \citet{guiglion16} using the MARCS model
atmospheres. In the case of Giraffe, where the lithium line is blended with
the nearby \ion{Fe}{i} line at 6707.4~\AA{} and only the total Li+Fe EW can
be measured, a correction for the Fe contribution (measured on the same
spectral grid of the curves of growth) was applied to derive the Li-only EW
before the abundance was computed. For M-type stars, in which the true
continuum is masked by the molecular bands and Li is heavily blended with
other components, the procedure described above cannot be applied, and only
a pseudo-EW can be measured: for these stars, we derived the curves of
growth and the pseudo-EWs in a consistent way by integrating over a
predefined interval that increased linearly with rotation rate above
20~km~s$^{-1}$. A more detailed description of the method is provided in
Franciosini et al. (in prep.). Uncertainties on the Li abundances were
derived taking the uncertainties on the EW measures and on the stellar
parameters into account. The dominant contribution to the abundance
uncertainties comes from the EWs themselves and from $T_\mathrm{eff}$.

We complemented the GES data with astrometry and photometry from {\em Gaia}
EDR3 \citep{gaia16,gaia21}. To search for possible additional members, in
particular at the bright ends of the cluster sequences that are poorly
sampled by GES, we extracted the {\em Gaia} data over slightly larger
regions than the GES fields, using a radius of 1~deg (or 1.6~deg for
NGC~2451) around the cluster centres, consistent with the cluster radii. 

The {\em Gaia} and GES catalogues for each cluster were cross-matched in
TOPCAT\footnote{http://www.starlink.ac.uk/topcat/} using a 2\arcsec\ search
radius to account for possible significant motions between the epoch of the
2MASS catalogue used for the GES observations and {\em Gaia} EDR3. We found
{\em Gaia} counterparts for all GES targets, except for seven objects in
Gamma~Vel. The remaining {\em Gaia} objects not present in GES were then
added to the catalogue to produce a combined GES+{\em Gaia} dataset for each
cluster. In the following, we use the terms `GES sample' and `{\em Gaia}
sample' to distinguish the stars observed by GES from those that are present
only in {\em Gaia}.

We removed from each combined catalogue all stars classified as secure
double-lined spectroscopic binaries in the GES catalogue. We also removed
those for which both the GES RV and {\em Gaia} astrometry are lacking
because no accurate membership can be derived for them. 

We defined an astrometric quality flag to identify stars with less reliable
astrometry. Following \citet{lindegren21} and \citet{fabricius21}, we used
the renormalised unit weight factor ({\tt ruwe}) to identify possible
spurious solutions due to potential doubles and flagged the sources with
{\tt ruwe} $> 1.4$. We additionally flagged also stars with a relative error
on the parallax greater than 20\% ({\tt parallax\_over\_error} $<5$). We
verified that this value is the best compromise for removing the bulk of
background contaminants without excluding any potential faint cluster
member. All objects with flagged astrometry and no GES RV were then removed
from the samples. For the stars in the GES sample with flagged astrometry,
only the RV was used for the membership selection.

Following the prescriptions in \citet{riello21}, we corrected the {\em Gaia}
magnitudes of bright stars for saturation effects as well as for the
systematic effects in the $G$-band magnitudes of stars fainter than $G=13$
with a six-parameter astrometric solution\footnote{%
These are sources without prior reliable colour information, for which an
additional pseudo-colour parameter was required for the astrometric solution
\citep[see][]{gaia21}.}. %
{\em Gaia} photometry may be affected by a flux excess in the
$G_\mathrm{BP}$ and $G_\mathrm{RP}$ bands for faint sources or in crowded
regions. We therefore also added a photometric flag to identify stars with
reliable {\em Gaia} photometry by applying the criterion defined by
\citet{riello21}, that is, $\vert C^*\vert < N \sigma_{C^*}(G)$, where $C^*$
is the flux excess factor corrected for the colour dependence,
$\sigma_{C^*}(G)$ is its scatter as a function of $G$,  and $N$ is the
chosen $\sigma$ threshold. Based on the distribution of $C^*$ for the
sources selected in Sect.~\ref{sec:rvastr}, we adopted $N=3$. Moreover, we
selected as good only stars with an error smaller than 5\% in the
$G_\mathrm{BP}$ and $G_\mathrm{RP}$ magnitudes. Stars that failed to fulfill
these criteria were not included in the cluster CMDs.

\begin{figure*}
\resizebox{\hsize}{!}{\includegraphics{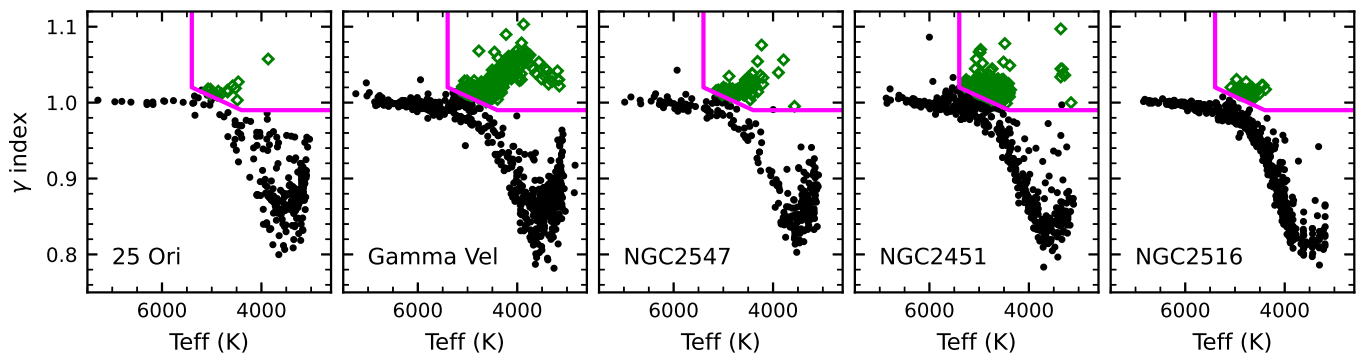}}
\caption{Member selection in the $\gamma$ vs $T_\mathrm{eff}$ plane for
the five clusters in our sample. The magenta lines indicate the
exclusion region we adopted for the selection. Stars marked with green
diamonds have been excluded as contaminants.}
\label{fig:gamma_sel}
\end{figure*}

\begin{figure*}
\resizebox{\hsize}{!}{\includegraphics{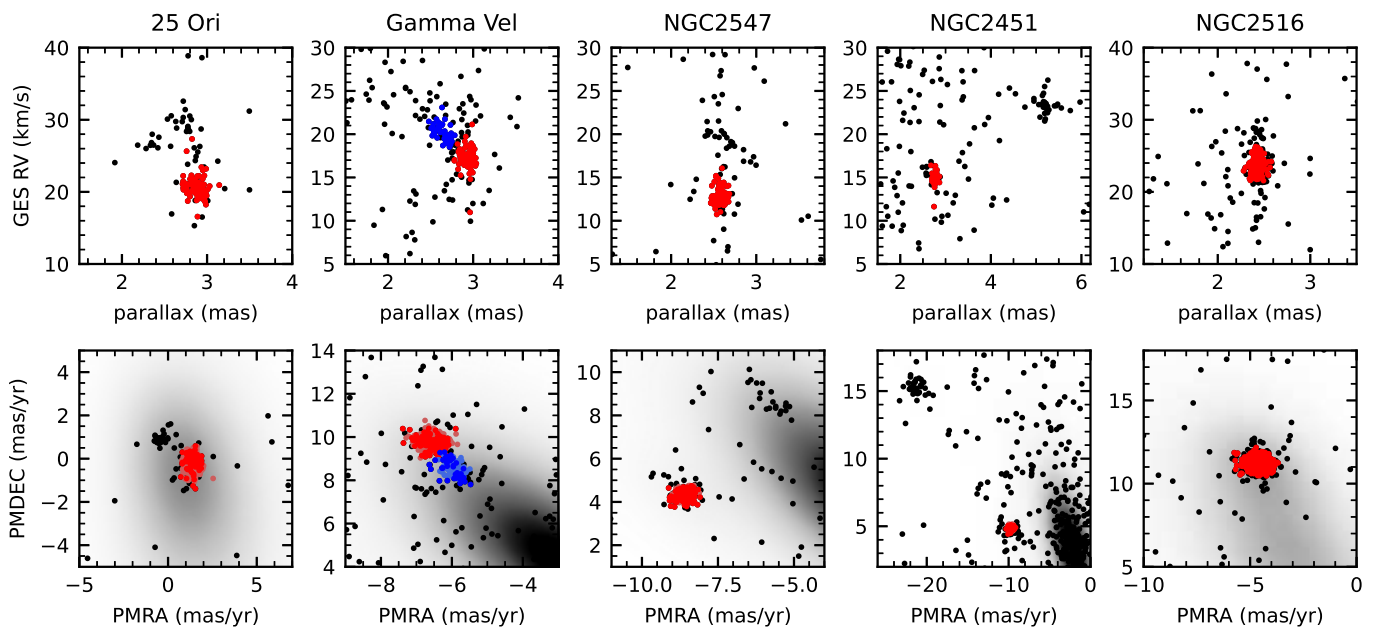}}
\caption{GES RV versus parallax ({\it top panels}) and proper motion
diagrams ({\it bottom panels}) for the GES samples for the five clusters
after the gravity selection (black dots). In the proper motion diagrams we
also plot the density map of the full {\em Gaia} samples on a grey scale.
The final samples of selected GES and {\em Gaia} members are overplotted
as red and light red dots, respectively, for 25~Ori, Gamma~Vel~A,
NGC~2451\,B, NGC~2547 and NGC~2516, and as blue and light blue dots,
respectively, for Gamma~Vel~B.
}
\label{fig:rvastr}
\end{figure*}

\section{Membership analysis}
\label{sec:membership}

The selection of cluster members was performed using a consistent approach
for all clusters to ensure sample homogeneity. To avoid introducing any
possible biases when comparing the observational CMDs and the lithium
patterns of the different clusters with theoretical models, our membership
procedure was based only on gravity, RVs, and {\em Gaia} astrometry as
primary selection criteria. Photometry and lithium were inspected only at
the end to exclude a few possible residual outliers.

\subsection{Gravity}
\label{sec:gravity}

The unbiased target selection strategy adopted in the GES (see
Sect.~\ref{sec:data}) implies a high level of contamination of the observed
sample by background giants and older field stars. However, background
giants can be easily identified on the basis of gravity. This allowed us to
remove them from the sample.

Measurements of $\log g$ are available for all UVES spectra, but only for a
fraction of stars observed with Giraffe. However, as a proxy for gravity, we
used the spectroscopic index $\gamma$, defined by \citet{damiani14} for
HR15N spectra. When $\gamma$ is plotted as a function of temperature or
colour, giant stars occupy a well-defined cloud at $\gamma \ga 1$ that is
clearly separated from the cluster locus, as shown in
Fig.~\ref{fig:gamma_sel} \citep[see also][]{bravi18}. For consistency, we
adopted a common selection that was valid for all clusters, although this is
less efficient for the older ones: in particular, we excluded all stars
falling within the region delimited by the magenta lines in
Fig.~\ref{fig:gamma_sel}, taking errors into account (i.e. if the entire
error bar is included in the region). Some of the clusters include a few
ambiguous cases that are located very close to the boundary of the region.
Their RV and/or astrometry is inconsistent with membership, however, and if
they were not discarded by the above selection, they were rejected by the
analysis described in the next section.

For stars observed only with UVES, we excluded all objects with $\log g <
3.5$~dex and 4000~K $< T_\mathrm{eff} <$ 5500~K. This selection was
also applied to stars observed with Giraffe with an available $\log g$
value to reject a few residual giants that could not be identified using
the $\gamma$ index.

\begin{table*}[t]
\centering
\caption{Maximum likelihood best-fit parameters of the RV and astrometry
distributions of the sample clusters\label{tab:ml_fit}}
\tiny
\begin{tabular}{lcccccccc}
\hline\hline\noalign{\smallskip}
Cluster& RV& $\sigma_\mathrm{RV}$& $\varpi$& $\sigma_\varpi$& $\mu_{\alpha*}$&
$\sigma_{\mu_\alpha*}$& $\mu_\delta$& $\sigma_{\mu_\delta}$ \\
 & (km~s$^{-1}$)& (km~s$^{-1}$)& (mas)& (mas)& (mas~yr$^{-1}$)&
(mas~yr$^{-1}$)& (mas~yr$^{-1}$)& (mas~yr$^{-1}$) \\
\noalign{\smallskip}\hline\noalign{\smallskip}
25~Ori     & $20.49\pm 0.08$& $0.83\pm 0.07$&
  $2.881\pm 0.004$& $0.051\pm 0.004$&  $\>\>1.347\pm 0.015$& $0.282\pm 0.013$&  
  $-0.178\pm 0.017\>\>$& $0.348\pm 0.014$ \\
Gamma~Vel~A& $17.13\pm 0.08$& $0.70\pm 0.07$& 
  $2.915\pm 0.005$& $0.058\pm 0.004$&  $-6.502\pm 0.016$& $0.278\pm 0.012$&  
  $9.706\pm 0.018$& $0.302\pm 0.014$ \\
Gamma~Vel~B& $20.19\pm 0.17$& $1.10\pm 0.14$& 
  $2.645\pm 0.008$& $0.077\pm 0.007$&  $-5.983\pm 0.021$& $0.265\pm 0.016$&  
  $8.609\pm 0.035$& $0.434\pm 0.024$ \\
NGC~2547    & $12.76\pm 0.07$& $0.74\pm 0.06$& 
  $2.588\pm 0.002$& $0.019\pm 0.006$&  $-8.567\pm 0.014$& $0.239\pm 0.011$&  
   $4.334\pm 0.014$& $0.231\pm 0.011$ \\
NGC~2451\,B & $15.02\pm 0.10$& $0.47\pm 0.10$& 
  $2.766\pm 0.004$& $0.037\pm 0.014$&  $-9.686\pm 0.019$& $0.258\pm 0.021$& 
  $4.854\pm 0.014$& $0.183\pm 0.012$ \\
NGC~2516    & $23.82\pm 0.06$& $0.96\pm 0.05$& 
  $2.429\pm 0.001$& $0.025\pm 0.003$&  $-4.622\pm 0.012$& $0.438\pm 0.010$& 
 $11.239\pm 0.010\,\>$& $0.365\pm 0.009$ \\
\hline
\end{tabular}
\end{table*}

\subsection{Radial velocity and astrometry}
\label{sec:rvastr}

Figure~\ref{fig:rvastr} shows the RV versus parallax and the proper motion
diagrams of the GES samples of the five clusters after the gravity
selection. The 25~Ori cluster is the group of stars at
RV$\sim\,$20~km~s$^{-1}$, $\varpi\sim\,$2.9~mas, and
$(\mu_\alpha^*,\mu_\delta)\sim (1.5,-0.5)$. The other smaller group at
$(\mu_\alpha^*,\mu_\delta)\sim (0,1)$ is the secondary population
\citep{zari19}, which corresponds to the group of stars around
RV\,$\sim$\,30~km~s$^{-1}$. The two populations of Gamma~Vel
\citep{jeffries14} can be clearly identified in both diagrams. Gamma~Vel~A
is concentrated at RV$\sim\,$17~km~s$^{-1}$, $\varpi\sim\,$2.9~mas, and
$(\mu_\alpha^*,\mu_\delta)\sim (-6.5,9.7)$, and Gamma~Vel~B is distributed
along a more elongated structure \citep[see][]{francio18}. In the case of
NGC~2547, the B population from \citet{sacco15} can be seen as a distinct,
small group of stars at RV$\sim\,$20~km~s$^{-1}$ and
$(\mu_\alpha^*,\mu_\delta)\sim (-5.5,8.5)$.
The diagrams for NGC~2451 show that the two clusters are very well
separated in all parameters: NGC~2451\,B is the group with lower RV,
parallax, and total proper motion. NGC~2516 shows a single peak without
evidence for additional populations. 

Figure~\ref{fig:rvastr} clearly shows that the samples still contain a
significant number of field contaminants. This is particularly true for
the {\em Gaia} samples, in which the number of background sources is
extremely large. To remove the bulk of contaminants and reduce the samples
to manageable sizes, we used the GES samples to derive first-guess average
centroids and standard deviations of the parallaxes and proper motions for
each population, applying a 3$\sigma$ clipping. We then excluded all stars
with reliable astrometry from the full samples that lie at more than
5$\sigma$ from these centroids, taking the individual errors into account.
That is, we retained only stars satisfying the
following relation:
\begin{equation}
\frac{(\varpi_i-\varpi_\circ)^2}{\sigma_{\varpi,i}^2+\sigma_{\varpi\circ}^2}
\, + \, 
\frac{(\mu_{\alpha*,i}-\mu_{\alpha*\circ})^2}{\sigma_{{\mu_{\alpha*}},i}^2+
\sigma_{\mu_{\alpha*}\circ}^2} \, + \,
\frac{(\mu_{\delta,i}-\mu_{\delta\circ})^2}{\sigma_{{\mu_\delta},i}^2+
\sigma_{\mu_\delta\circ}^2} \le 25 
\end{equation}
for at least one cluster population. Adopting a higher threshold would
significantly increase the number of field objects, especially for the
younger clusters. This would complicate the fit described below and increase the
risk of including contaminants in our final samples. For NGC~2547, we
considered only the main cluster and discarded the secondary population since
it cannot be constrained by the fit. For 25~Ori we kept both populations for
the fit, although only members of the main cluster were selected
afterwards. Because the two NGC~2451 clusters are well separated in all
parameters, it might be possible in principle to select only stars around
the centroid of cluster B, but the GES sample contains a significant number
of stars lacking reliable astrometry (to which the above selection therefore
cannot be applied) but with an RV that is consistent with that of
NGC~2451\,A. For this reason, we preferred to fit the data for both clusters
and then selected the members of NGC~2451\,B afterwards.

For each cluster, we performed a simultaneous fit of the distributions of
RVs, parallaxes, and proper motions for the reduced datasets using a
maximum likelihood approach and assuming that the total probability
distribution is described by the sum of one or two cluster populations
plus the field. To do this, we extended the method described by
\citet{lindegren00} that was used by \citet{francio18} and \citet{roccat18} and
included the RV as a fourth parameter. A detailed description is given in
Appendix~\ref{sec:mlfit}. The resulting best-fit values for the cluster
components are given in Table~\ref{tab:ml_fit}. Individual membership
probabilities $P_i$ for each population were derived using the standard
method, that is, dividing the probability of belonging to the given population
by the total probability. For the following analysis, we selected only the
highest-probability members of each cluster population with $P_i > 0.96$.

\subsubsection{Parallax zero-point correction}
\label{sec:zp}

\citet{lindegren21plx} investigated the systematic bias affecting {\em
Gaia} EDR3 parallaxes, which shows a complex dependence on magnitude,
colour, and ecliptic latitude, and provided a tentative correction for the
parallax zero-point. \citet{fabricius21} showed that applying this
correction to distant clusters improves their parallax distribution
reducing the internal dispersion. Hence, we tested the effect of the bias
correction on our results. 

The zero-point for each star was computed using the Python code provided
on the {\em Gaia} web pages\footnote{%
https://www.cosmos.esa.int/web/gaia/edr3-code}, %
and the whole analysis of Sect.~\ref{sec:rvastr} was repeated using the
corrected parallaxes. For all clusters, we found no significant difference
in the best-fit parameters, except for the central parallax, or in the list
of selected members. The mean parallax is increased by $0.026-0.039$~mas,
depending on the cluster. In the following, we therefore adopt the results
of Sect.~\ref{sec:rvastr}, although we discuss the effect of this parallax
bias on the model comparison.

\begin{figure*}
\centering
\resizebox{0.84\hsize}{!}{\includegraphics{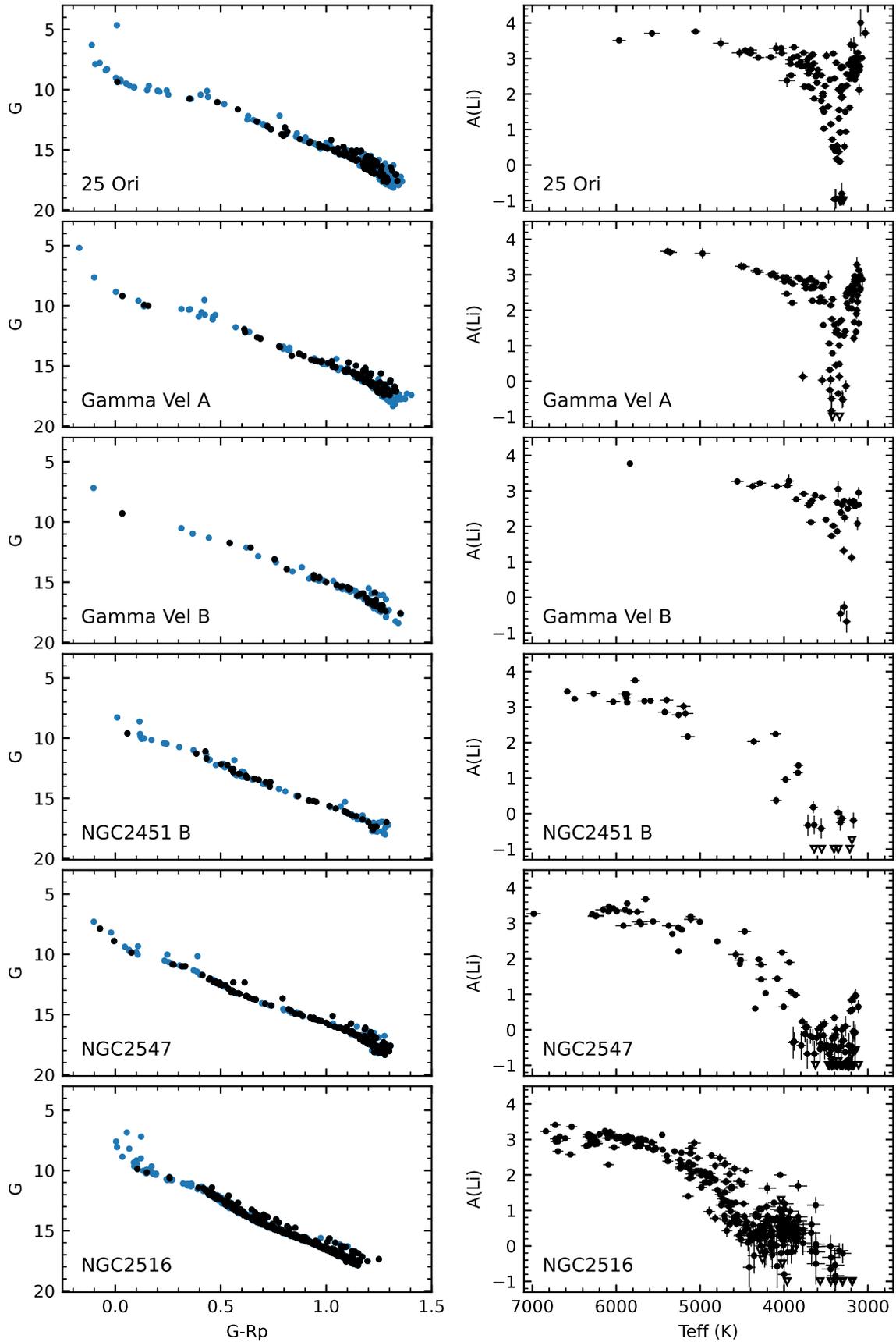}}
\caption{
CMDs ({\it left panels}) and lithium distribution as a function of
$T_\mathrm{eff}$ ({\it right panels}) of the final samples of selected
members for each cluster. The GES and {\em Gaia} members are indicated by
black and blue dots, respectively. In the left panels, only stars with good
photometry are plotted. The errors on photometry are smaller than the symbol
size. Downward triangles in the lithium diagrams indicate upper limits.}
\label{fig:finalmem}
\end{figure*}

\subsection{Photometry and lithium}
\label{sec:phli}

As a final step, we examined the photometry and the lithium abundances to
identify possible residual contaminants in our samples. The sequences of the
selected high-probability members with reliable photometry are very clean in
the $G$ versus $G-G_\mathrm{RP}$ CMD. However, we found one star in
Gamma~Vel~B and one in NGC~2451\,B that lay more than 1~mag below the
cluster sequence. These objects were discarded from the samples. Similarly,
we discarded one star in 25~Ori and one in NGC~2547, both with
$T_\mathrm{eff}\sim\,$5300~K and lithium $\sim$\,2~dex below the lower
envelope of the cluster sequences in the $A$(Li) vs $T_\mathrm{eff}$
diagrams, since these are likely to be non-members. 

\subsection{Final members}
\label{sec:finalmem}

The selection procedure described above allowed us to obtain very clean
lists of secure members, which are fundamental for a precise comparison
with theoretical models. The final membership lists contain 442 members of
25~Ori (137 GES), 325 members of Gamma~Vel~A (118 GES), 128 of Gamma~Vel~B
(44 GES), 255 of NGC~2547 (146 GES), 125 members of NGC~2451\,B (42 GES),
and 536 of NGC~2516 (363 GES). Their properties are listed in
Tables~\ref{tab:25ori}-B.6. 

The CMDs and lithium patterns of the final members are plotted in
Fig.~\ref{fig:finalmem}. All clusters show very clean sequences, although
some scatter is still present, especially at the lowest masses. For some
of them, the binary sequences are also well distinguished. The bright end
of the NGC~2516 sequence shows a significant spread. This might be
indicative of an extended main-sequence turn-off, which has recently been
observed in Galactic open clusters of different ages \citep[e.g.][and
references therein]{bastian18,marino18,li19}, and is generally attributed to
a spread in rotation rates. Unfortunately, we do not have information on
rotation for these stars that could confirm this hypothesis.

The right panels of Fig.~\ref{fig:finalmem} show the evolution of the
lithium abundances with age. The lithium depletion pattern of 25~Ori is very
similar to the pattern observed in Gamma~Vel~A and B \citep{jeffries14},
with a continuous range of lithium abundances for stars between $\sim$\,3200
and $\sim$\,3600~K, from nearly undepleted stars to fully depleted ones. To
our knowledge, this is the first determination of the lithium depletion
pattern in 25~Ori. The similarity with Gamma~Vel suggests that these
clusters have similar ages and that this particular depletion pattern is a
common characteristic of young clusters at this age. At the age of
NGC~2451\,B and NGC~2547, M-type stars are generally fully depleted,
although some of them still retain some amount of lithium; we also note the
start of the lithium increase at the LDB that is clearly visible in
NGC~2547. At the age of NGC~2516, lithium depletion has proceeded to higher
masses, although a strong scatter is present. Several stars retain a
significant amount of lithium even below 4000~K.

The younger clusters include some stars, between $\sim$\,4800 and
6000~K, whose lithium abundance is significantly higher than the initial
cosmic abundance $A\mathrm{(Li)}\sim 3.3$ \citep[e.g.][]{martin94,romano21}.
This is a signature of non-local thermodynamic equilibrium (NLTE) effects:
if the three-dimensional NLTE corrections by \citet{wang21} were applied,
these abundances would be reduced by $\sim$\,0.3--0.4~dex, bringing them in
agreement with the cosmic value. At higher temperatures, the corrections are
not significant, being $<0.1$~dex. However, the \citet{wang21} grid is
limited to $T_\mathrm{eff}\ge 4000$~K and does not cover the full range of
temperatures in our samples. For this reason, the corrections were not
applied to our data. The high lithium abundance of the two stars at 3000~K
in 25~Ori is instead likely caused by an underestimated gravity due to the
low signal-to-noise ratio of the spectra: their reported $\log g$ is
$\sim$\,3.2--3.4~dex, but their $\gamma$ index and spectra indicate that
they are dwarfs. A more reasonable value of $\log g=4.0$~dex would lower
their abundances to $\sim$\,3.3.

\begin{figure*}[!ht]
\centering
\resizebox{\hsize}{!}{%
\includegraphics[width=0.32\hsize]{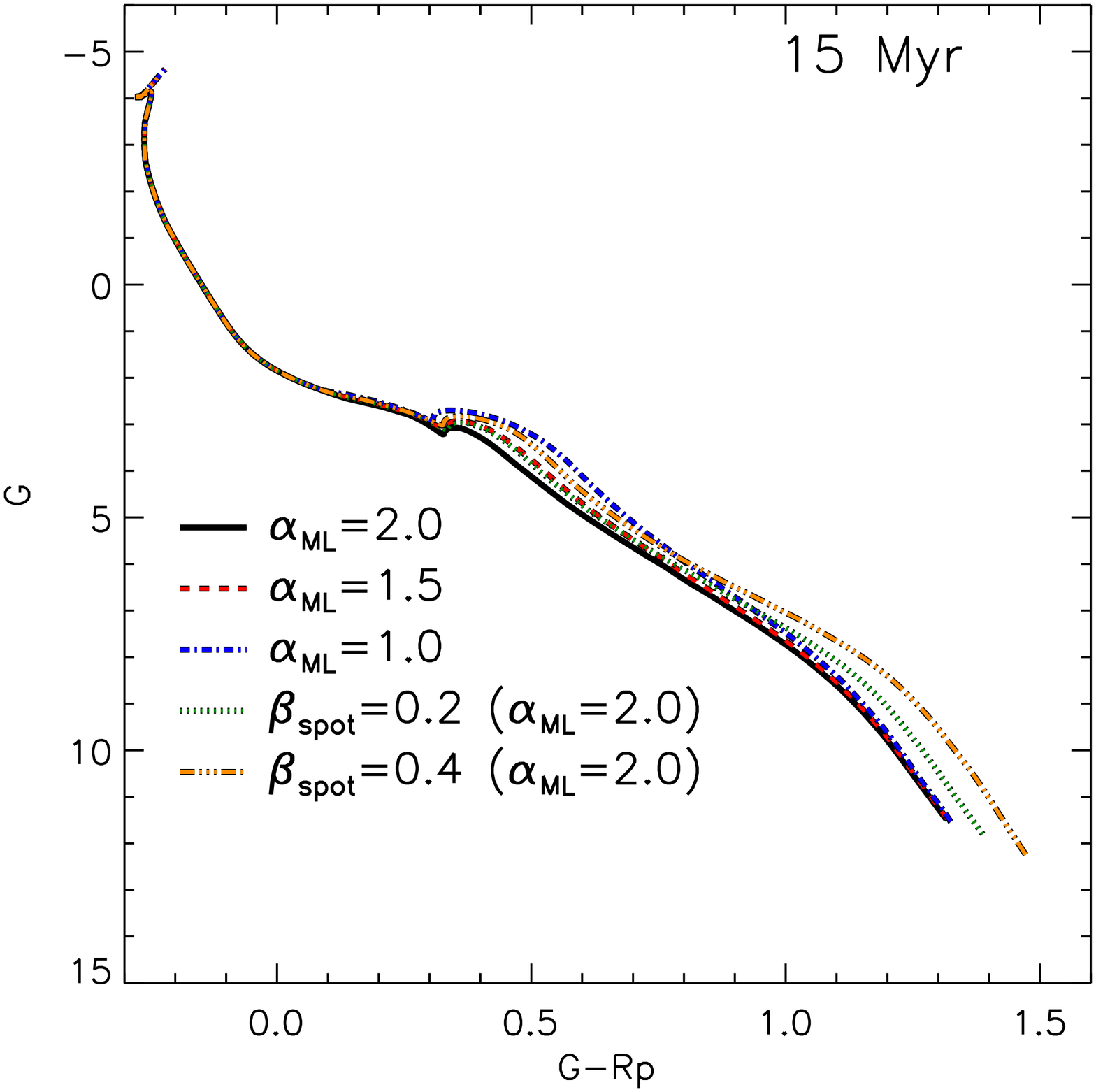}
\includegraphics[width=0.32\hsize]{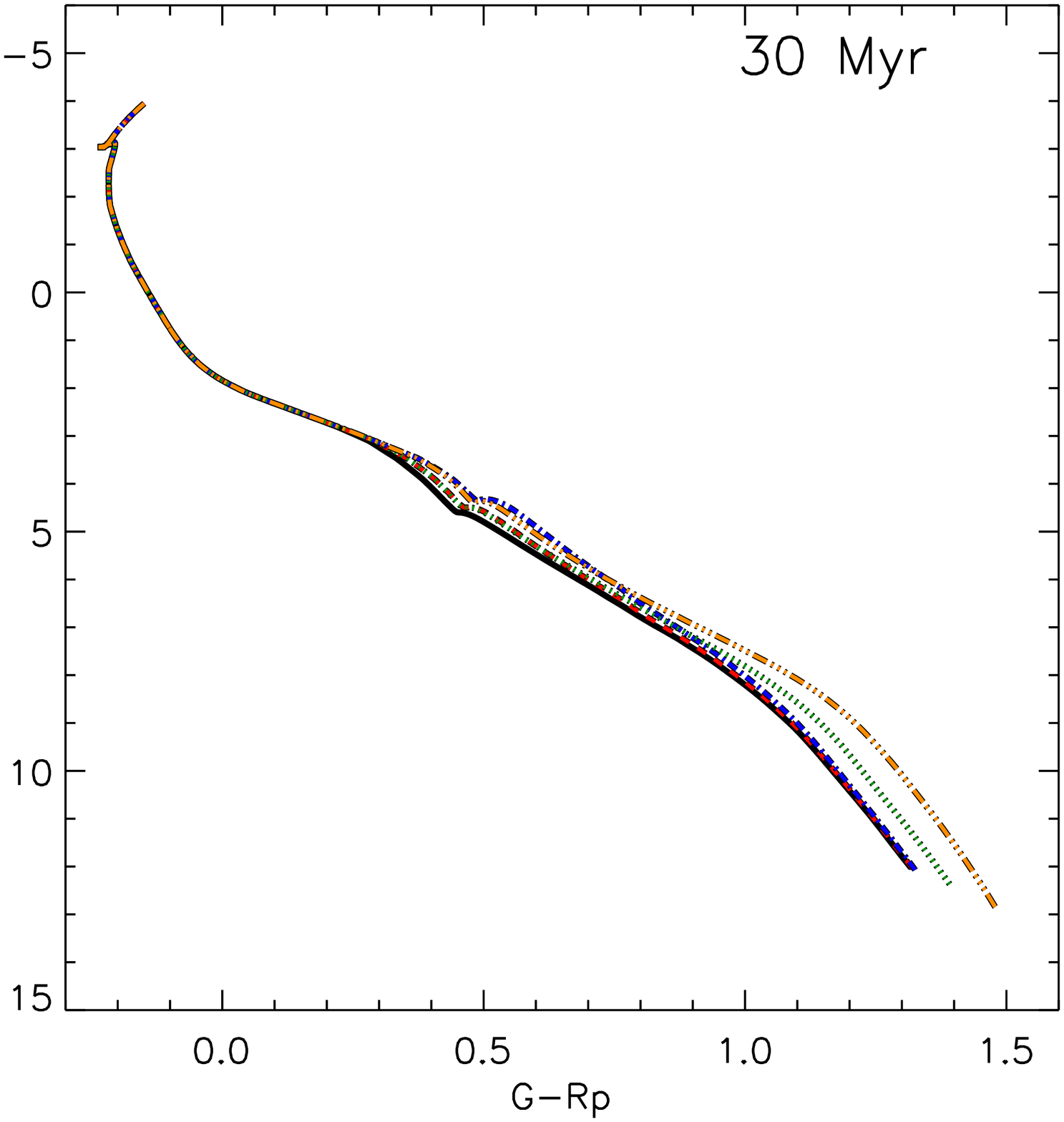}
\includegraphics[width=0.32\hsize]{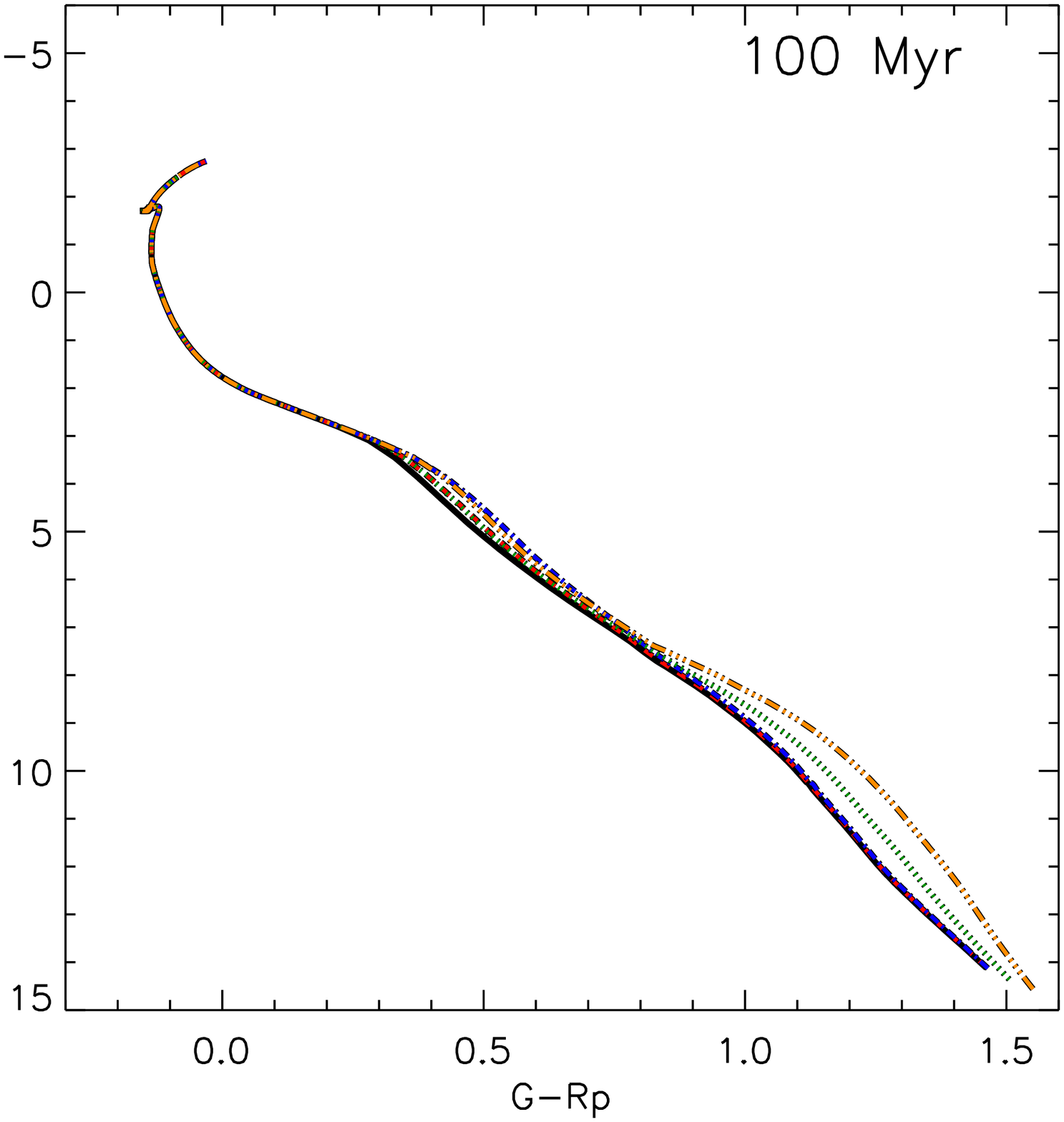}
}
\resizebox{\hsize}{!}{%
\includegraphics[width=0.31\hsize]{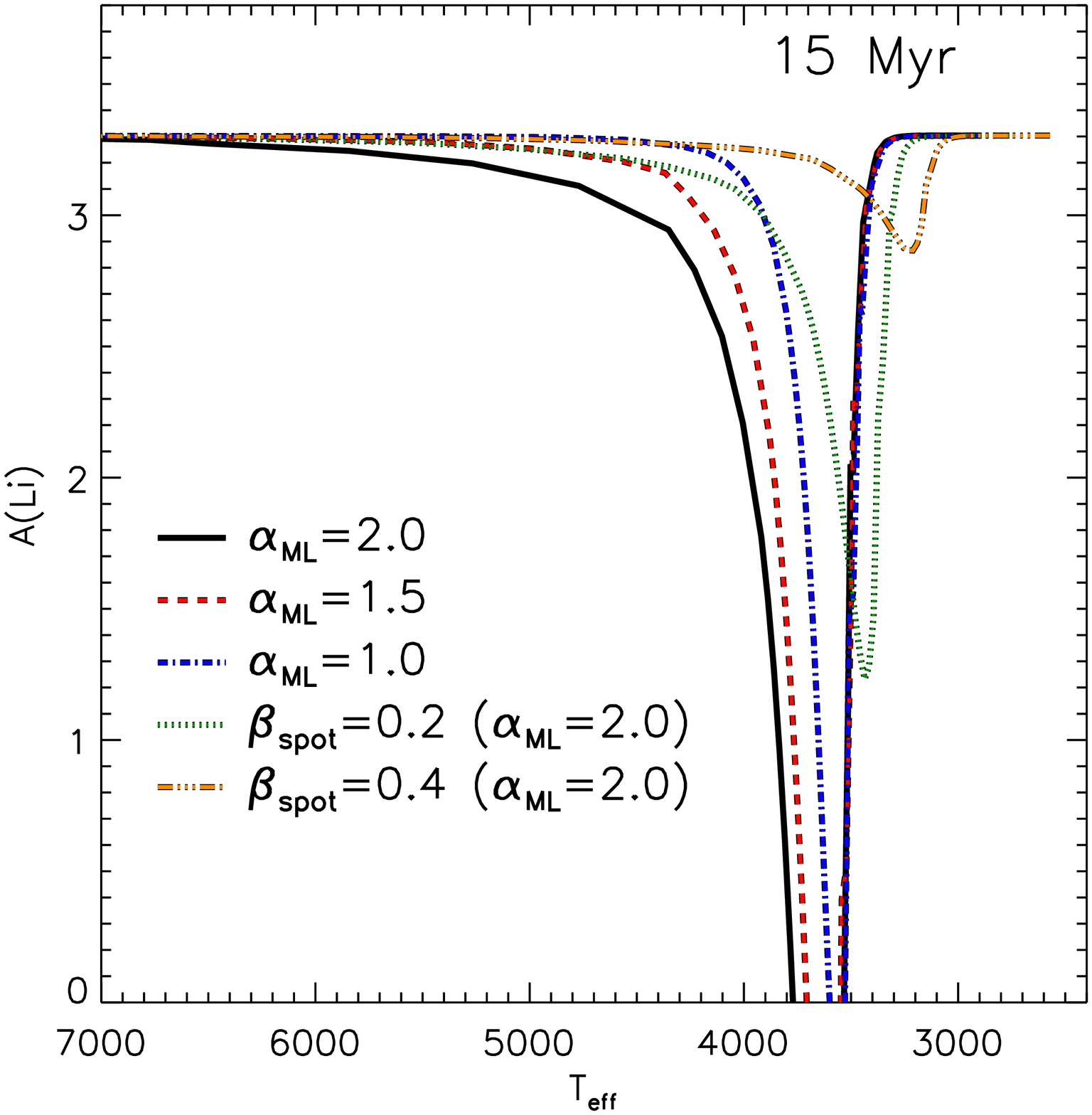}
\includegraphics[width=0.31\hsize]{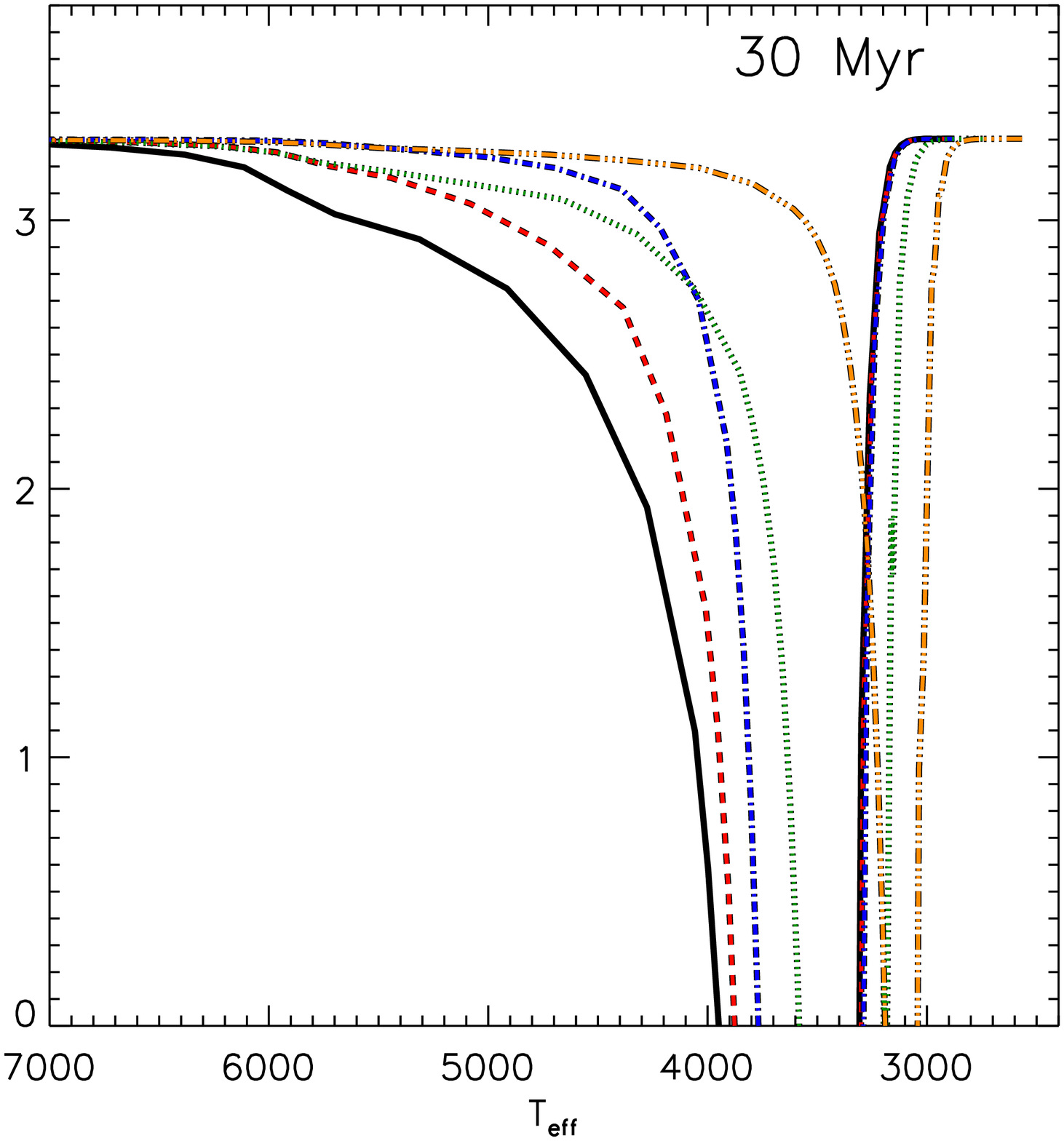}
\includegraphics[width=0.31\hsize]{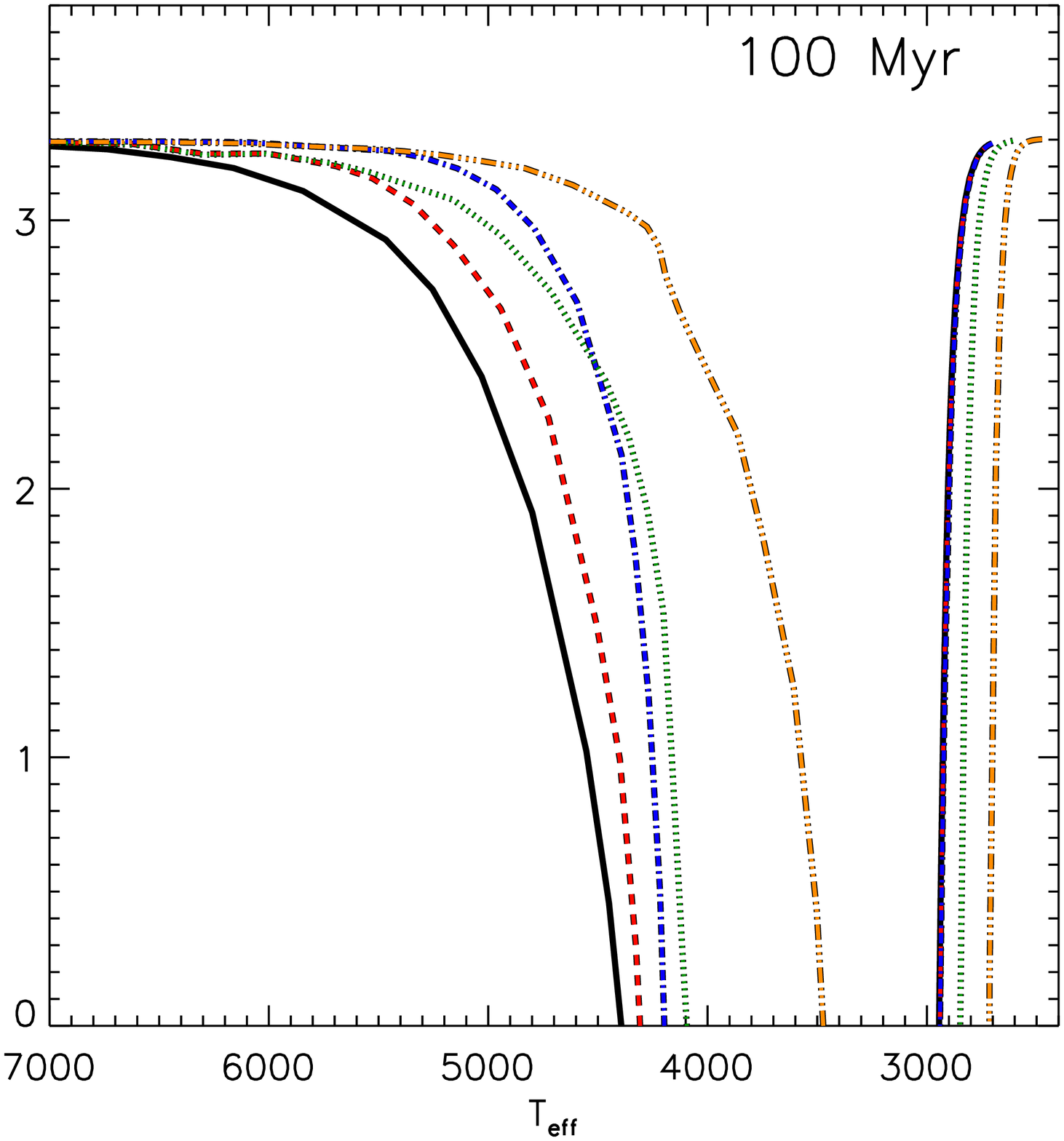}
}
\caption{Effect of adopting different values of $\alpha_\mathrm{ML}$
and of different spot coverage fractions on the models at 15~Myr ({\it left
panels}), 30~Myr ({\it middle panels}), and 100~Myr ({\it right panels}).
The top row shows the effect on the CMD, and the bottom row presents the
effect on the surface lithium abundance as a function of $T_\mathrm{eff}$.
The standard reference model with $\alpha_\mathrm{ML}=2$ and no spots is
plotted as the solid black line. The dashed red and dot-dashed blue lines
show the models without spots and $\alpha_\mathrm{ML}=1.5$ and 1.0,
respectively. The models with $\alpha_\mathrm{ML}=2$ and
$\beta_\mathrm{spot}=0.2$ and 0.4 are indicated with the dotted green and
three-dot-dashed orange lines, respectively.}
\label{fig:iso_ml_beta}
\end{figure*}

\section{Comparison with evolutionary models}
\label{sec:model_comp}

\subsection{Models}
\label{sec:models}

The models were calculated with the Pisa evolutionary code
\citep{tognelli11,dellomodarme12}. The main input physics and
parameters are described in detail in
\citet{tognelli15a,tognelli15b,tognelli18}. Here we only point out some 
relevant characteristics and some improvements with respect to the quoted 
works. 

The adopted radiative opacity tables are OPAL~2005%
\footnote{http://opalopacity.llnl.gov/new.html} 
\citep{iglesias96} for the stellar interiors and \citet{ferguson05}%
\footnote{http://webs.wichita.edu/physics/opacity} 
for the most external regions, both computed for the \citet{asplund09}
solar mixture. The equation of state is the OPAL~2006 \citep{rogers02},
consistent with the adopted opacities, extended to the low-mass star
regime ($M\la 0.2\,M_\sun$) by the \citet{saumon95} 
equation of state. 

Convective transport in super-adiabatic regions was treated following the
mixing length formalism by \citet{bohm58}, where the convection efficiency
is expressed in terms of the mixing length scale $\ell_\mathrm{ML} =
\alpha_\mathrm{ML}\,H_\mathrm{P}$, where $\alpha_\mathrm{ML}$  is the
mixing length parameter and $H_\mathrm{P}$ is the pressure scale height.
As reference value we adopted our solar-calibrated%
\footnote{Our standard solar model was computed using an iterative
procedure in which the initial helium abundance, metallicity, and mixing
length parameter were adjusted to reproduce the observed radius, luminosity,
and photospheric $(Z/X)$ in a 1~$M_\sun$ model at the solar age within a
given numerical tolerance ($< 10^{-4}$).} %
parameter $\alpha_\mathrm{ML}=2.0$. This value was shown to reproduce the
CMD of the Hyades cluster well \citep{tognelli21}. In addition, to test the
effect of a reduced convection efficiency on the models, we also considered
the two cases $\alpha_\mathrm{ML}=1.0$ and 1.5.

The most relevant nuclear reactions for our analysis are the
hydrogen-burning reactions, that is, the pp-chain and the CNO-cycle. Their
reaction rates are from \citet{adelberger11}, except for p(p,$e^+\nu)^2$H,
which is from \citet{marcucci13} and \citet{tognelli15a},
$^2$H(p,$\gamma)^3$He, which is from \citet{descouvement04},
$^7$Li(p,$\alpha)\,\alpha$, which is from \citet{lamia12}, and
p($^{14}$N,$\gamma)^{15}$O, which is from \citet{imbriani05}. 

We obtained the outer boundary conditions from the detailed non-grey
BT-Settl atmospheric structures%
\footnote{https://phoenix.ens-lyon.fr/Grids/BT-Settl/AGSS2009/}
by \citet{allard11} for $T_\mathrm{eff} \la 10\,000$~K and from the
\citet{castelli03} structures for higher $T_\mathrm{eff}$.
We also included the possibility in the models of having a partial coverage
of the stellar surface by starspots, which block part of the outgoing
radiation. To model the spots, we used a formalism similar to that described
in \citet{somers15}. We treated the spots as a pure surface effect, assuming
that they do not extend below the stellar surface, that is, that the spot
scale height is much smaller than the pressure scale height. The total
luminosity at the stellar surface is given by the following expression:
\begin{eqnarray}
L_\star& =& \sigma_\mathrm{SB} \left(\mathcal{A}_\mathrm{unspotted} 
    T_\mathrm{eff}^4 + \mathcal{A}_\mathrm{spotted} T_\mathrm{spot}^4\right) 
    \>=\nonumber\\
  &=& \sigma_\mathrm{SB} \mathcal{A}_\star T_\mathrm{eff}^4\,\left( 
    \frac{\mathcal{A}_\mathrm{unspotted}}{\mathcal{A}_\star} + 
    \frac{\mathcal{A}_\mathrm{spotted}}{\mathcal{A}_\star} 
    \frac{T_\mathrm{spot}^4 }{T_\mathrm{eff}^4}\right)\>=\nonumber\\
  &=& \sigma_\mathrm{SB} \mathcal{A}_\star T_\mathrm{eff}^4 
    \,\left(1-\beta_\mathrm{spot}\right)\,,
\end{eqnarray}
where $\beta_\mathrm{spot}$ is an effective spot coverage fraction, which
takes the ratio of the spot and stellar surface
effective temperature and the real fraction of area covered with spots into account. This is defined as
\begin{equation}
\beta_\mathrm{spot} =
\frac{\mathcal{A}_\mathrm{spotted}}{\mathcal{A}_\star} 
  \,\left(1 - \frac{T_\mathrm{spot}^4}{T_\mathrm{eff}^4} \right)\,. 
\end{equation}
The parameter $\beta_\mathrm{spot}$ is useful when information about the
real spot coverage $\mathcal{A}_\mathrm{spotted} / \mathcal{A}_\star$ or on
$T_\mathrm{spot}/T_\mathrm{eff}$ is not well constrained. In these cases,
it is sufficient to specify only the effective spot coverage instead of
requiring two free parameters (namely $T_\mathrm{spot}/T_\mathrm{eff}$ and
$\mathcal{A}_\mathrm{spotted}/\mathcal{A}_\star$). For the typical values
of $T_\mathrm{spot}/T_\mathrm{eff}\sim\,0.7-0.9$ observed in active stars
\citep[e.g.][and references therein]{berdyugina05},
$\beta_\mathrm{spot}\sim\,0.76-0.34\, \mathcal{A}_\mathrm{spotted} /
\mathcal{A}_\star$. We analysed the effect of
two different values of effective spot coverage on the models, namely
$\beta_\mathrm{spot}=0.20$ and 0.40, which are consistent with those
suggested by recent observations of late-type stars
\citep[e.g.][]{fang16,gully17,morris19}.

For a star with a given luminosity and effective temperature, the radius of
a spotted model is $R_\mathrm{spotted} = R_\mathrm{unspotted} /
\sqrt{1-\beta_\mathrm{spot}}$, which is larger than the radius of the
corresponding unspotted model. The effective surface that can freely radiate
the energy is reduced by a factor $(1-\beta_\mathrm{spot}),$ and
consequently, the total surface of the star has to increase by the same
factor to keep the luminosity constant.

As mentioned in Sect.~\ref{sec:intro}, all the clusters present in our
sample are fully compatible with a solar chemical composition, that is,
[Fe/H]\,=\,0. Using the \citet{asplund09} relative metal abundances, a
helium-to-metal enrichment ratio $\Delta Y/\Delta Z= 2$
\citep{tognelli21}, and a primordial helium abundance $Y_p = 0.2485$
\citep{cyburt04,aver12,aver13}, we obtained an initial helium and global
metallicity $Y = 0.2740$ and $Z=0.0130$, respectively. Concerning light
elements, we set the mass fractional abundance of deuterium to
$X_D=2\times 10^{-5}$, which is representative of the present Galactic D abundance
\citep{linsky06,sembach10}, and the initial ${}^7$Li abundance to
reproduce that measured in stars with $T_\mathrm{eff}>6000$~K, where
lithium is still undepleted. For all cases, we adopted a value of
$A(\mathrm{Li})= 3.3$, which is consistent with the average initial
abundance of young open clusters derived by \citet{romano21} using GES
iDR6 data.

Evolutionary tracks were computed in the mass interval
[0.06,~10]~$M_\sun$ with a variable mass spacing. From these tracks, we
calculated the isochrones in the age interval [0.5,~200]~Myr. Theoretical
isochrones were converted into {\em Gaia}~EDR3 magnitudes using the filter
response curves given in \citet{riello21} and computing the bolometric
corrections using the MARCS~2008 atmospheres \citep{gustafsson08} for
$T_\mathrm{eff} \le 8000$~K and the \citet{castelli03} ones for higher
$T_\mathrm{eff}$. 

Figure~\ref{fig:iso_ml_beta} shows the comparison between the isochrones at
15, 30, and 100~Myr (representative of our cluster sample) obtained for the
standard set of models ($\alpha_\mathrm{ML}=2.0$) for the sets with a smaller
mixing length parameter ($\alpha_\mathrm{ML}=1.0$ and 1.5) and without spots,
and the sets with $\alpha_\mathrm{ML}=2.0$ and $\beta_\mathrm{spot} = 0.2$
and 0.4. The effects on the CMD are shown in the top panels of
Fig.~\ref{fig:iso_ml_beta}. As is well known, a variation in
$\alpha_\mathrm{ML}$ mainly affects the colours of stars with thick and
superadiabatic envelopes, therefore its effect depends on stellar mass.
Very low-mass stars are fully convective but almost adiabatic, and thus
they are not sensitive to the adopted $\alpha_\mathrm{ML}$. As the mass
increases, the outermost layers of the star become superadiabatic, but the
convective envelope becomes increasingly thinner as the mass increases. As a
result, the effect of varying $\alpha_\mathrm{ML}$ is maximum in an
intermediate mass interval. Conversely, the effect of spots increases as
the stellar mass decreases, producing significantly redder isochrones for
M-type stars: this happens because spots modify the outer boundary
conditions, which become increasingly more important as the mass decreases
\citep[see e.g.][]{tognelli11,tognelli18}.

The bottom panel of Fig.~\ref{fig:iso_ml_beta} shows the surface lithium
abundance as a function of the effective temperature for the selected
ages. The LDB is clearly visible in all panels. As is well known, as the age
increases, the LDB moves to lower temperatures and its luminosity decreases
\citep[see e.g.][]{tognelli15b}. The effect of $\alpha_\mathrm{ML}$ and
$\beta_\mathrm{spot}$ on the surface lithium abundance is more evident than
that on the CMD. By reducing $\alpha_\mathrm{ML}$ or increasing
$\beta_\mathrm{spot}$, the models become cooler and lithium is less
efficiently destroyed. 

The two parameters $\alpha_\mathrm{ML}$ and $\beta_\mathrm{spot}$ affect
the models in different ways. As mentioned before, a variation in
$\alpha_\mathrm{ML}$ changes the structure only in a limited mass
interval. For very low-mass stars (i.e. almost adiabatic ones, for
$T_\mathrm{eff} \la 3500$~K), the variation in $\alpha_\mathrm{ML}$ has a
negligible effect on  $T_\mathrm{eff}$ and on the lithium burning, which
results in no significant change of the LDB. On the other hand, for
higher masses (3500~K $\la T_\mathrm{eff} \la 6500$~K), the structure
and lithium burning are affected by a change in $\alpha_\mathrm{ML}$,
and this produces a shift in the left side of the lithium `chasm'
\citep{basri97} towards lower temperatures as $\alpha_\mathrm{ML}$
decreases. In contrast, a change in $\beta_\mathrm{spot}$ affects all stars
with $T_\mathrm{eff} \la 6500$~K and moves not only the hot side of the
chasm, but also moves the LDB slightly towards lower temperatures as
$\beta_\mathrm{spot}$ increases. Furthermore, the amount of lithium
depletion in the chasm significantly decreases as the spot coverage
increases at 15 Myr, and the LDB may not be clearly defined. The inclusion
of spots also affects the luminosity of the model at the LDB, and
consequently, its derived age. \citet{somers15} estimated an increase in LDB
age of about 13-15\% in the age interval 15-30~Myr, assuming a
(real) coverage fraction $f_\mathrm{spot}=0.5$%
\footnote{\citet{somers15} assumed a value of
$T_\mathrm{spot}/T_\mathrm{eff} = 0.8$ and a maximum value of
$f_\mathrm{spot} =
\mathcal{A}_\mathrm{spotted}/\mathcal{A}_\star=0.5$. Their models
with $f_\mathrm{spot} = 0.5$ correspond to our value of
$\beta_\mathrm{spot} \approx 0.3$.}. %
Our model grid was not computed with the aim to provide LDB models,
and it is not dense enough in mass to allow a precise identification of the
LDB at young ages, as was done in \citet{tognelli15b}. 
Nevertheless, we estimate that the age increases by approximately 8$-$10\%
at 15$-$30~Myr and $\sim\,$6\% at 100~Myr in the case of
$\beta_\mathrm{spot} = 0.2$, while the relative age increase is about
20$-$24\% at 15$-$30~Myr and 15\% at 100~Myr for $\beta_\mathrm{spot} =
0.4$. Similar variations were found by \citet{binks21} from their analysis
of the LDB in NGC~2232. 
Therefore, the very different effects produced by the two parameters
$\alpha_\mathrm{ML}$ and $\beta_\mathrm{spot}$ on the lithium pattern should
allow us at least in principle to distinguish which one of them is dominant.

\subsection{Determination of cluster reddening and age}
\label{sec:age}

The age and reddening $E(B-V)$ for each cluster were derived using the
Bayesian maximum likelihood method described in \citet{randich18} and
\citet{tognelli21} \citep[see also][]{hatzidimitriou19}. The analysis is
based on a star-by-star comparison of the observed {\em Gaia} EDR3
magnitudes with our theoretical isochrones. The age and $E(B-V)$ were
derived by constructing a two-dimensional likelihood: the likelihood peak
was used to obtain the best value while the uncertainty was evaluated by
marginalising the distribution and finding the regions in its wings that
account for 16\% of the total area. To perform the analysis, we used a
combination of all three {\em Gaia} magnitudes, considering in particular
the plane ($G-G_\mathrm{RP}$, $G_\mathrm{BP}$). We selected only stars
with an absolute $G_\mathrm{BP}$ magnitude lower than 7.5 (corresponding to
about 0.7--0.8~$M_\sun$ for ages older than 20--30~Myr). In this
region, models and observation attain a satisfactory agreement. At higher
magnitudes (in particular for $G-G_\mathrm{RP} \ga 0.9$), some problems
persist (not well understood) that prevent a good agreement
between models and data, such as difficulties in the modelling of synthetic
spectra at solar and supersolar metallicities, or in predicting the
radius/$T_\mathrm{eff}$ relation in fully convective PMS low-mass stars
\citep[see e.g.][]{kucinskas05,gagne18,tognelli21}. We therefore decided to
exclude this part of the data from the recovery to avoid introducing
an uncontrollable bias. We also excluded stars that lie on the
binary sequence or that are clearly outliers from the fit.

To evaluate the uncertainties on the results, we took the errors coming from
the photometry (mean $\Delta (G-G_\mathrm{RP})$ and $\Delta
G_\mathrm{BP}\sim\,$2$-$5~mmag) and the systematic parallax bias of
$\sim\,$0.03~mag derived in Sect.~\ref{sec:zp} into account. The impact of
the zero-point correction is small: it accounts for an uncertainty of about
3$-$5~mmag in $E(B-V)$ and an age increase of 1$-$2~Myr. 

If the stars have the same physical parameters and the photometric errors
are reliable, we would expect them to lie on the same isochrone. However,
the CMDs show that this is not the case (see Fig.~\ref{fig:finalmem}), as
in all clusters a scatter larger than the observable errors is present.
This could mean that some additional mechanisms, different from star to
star, might play a non-negligible role. Hence, we only aim to reproduce the
average characteristics of the stellar population in our comparison. 

The procedure was applied to different sets of models: (i) the reference
case with solar-calibrated mixing length ($\alpha_\mathrm{ML}=2.0$) without
spots, (ii) the sets with $\alpha_\mathrm{ML}=1.0$ and
$\alpha_\mathrm{ML}=1.5$ without spots, and (iii) the sets with
$\alpha_\mathrm{ML}=1.0$, 1.5, and 2.0 and $\beta_\mathrm{spot} =0.2$ and
0.4. 

\begin{table}
\centering
\caption{Best-fit model parameters for each cluster \label{tab:model_fit}}
\begin{tabular}{lcccc}
\hline\hline
Cluster & age  & $E(B-V)$& $\alpha_\mathrm{ML}$& $\beta_\mathrm{spot}$\\
        & (Myr)& (mag)  &                      &                      \\
\hline
\noalign{\smallskip}
Gamma~Vel~A& $18.0^{+1.5}_{-4.0}$& $0.062^{+0.006}_{-0.022}$& 1.0& 0.2\\[3pt]
Gamma~Vel~B& $21.0^{+3.5}_{-3.0}$& $0.088^{+0.006}_{-0.026}$& 1.0& 0.2\\[3pt]
25~Ori     & $19.0^{+1.5}_{-7.0}$& $0.065^{+0.023}_{-0.043}$& 1.0& 0.2\\[3pt]
NGC~2451\,B& $30.0^{+3.0}_{-5,0}$& $0.130^{+0.012}_{-0.031}$& 2.0& 0.0\\[3pt]
NGC~2547   & $35.0^{+4.0}_{-4.0}$& $0.106^{+0.028}_{-0.031}$& 2.0& 0.0\\[3pt]
NGC~2516   & $138^{+48}_{-42}$   & $0.154^{+0.026}_{-0.038}$& 2.0& 0.0\\[3pt]
\hline
\end{tabular}
\end{table}

\begin{figure*}
\centering
\resizebox{\hsize}{!}{%
\includegraphics{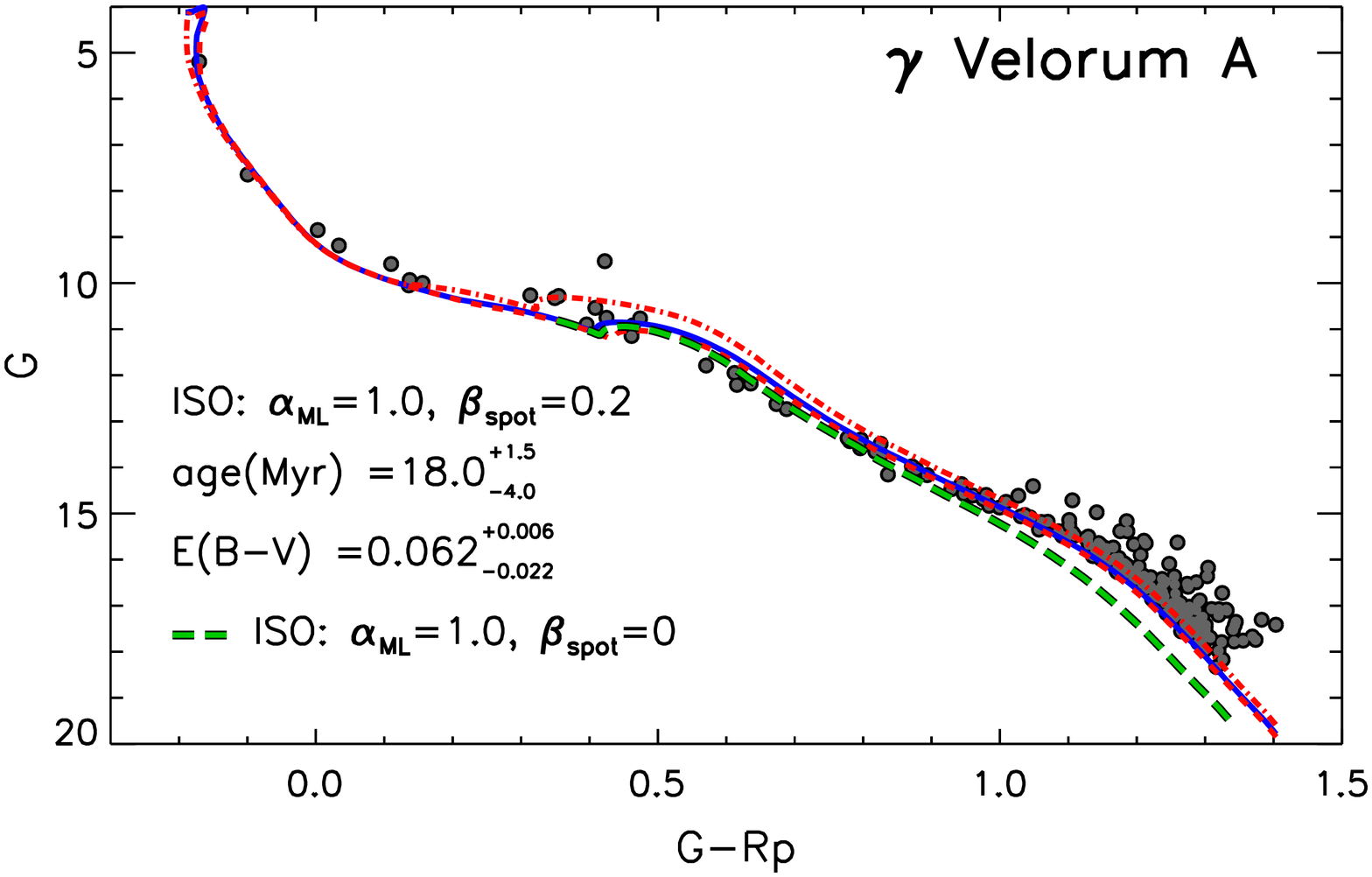}
\includegraphics{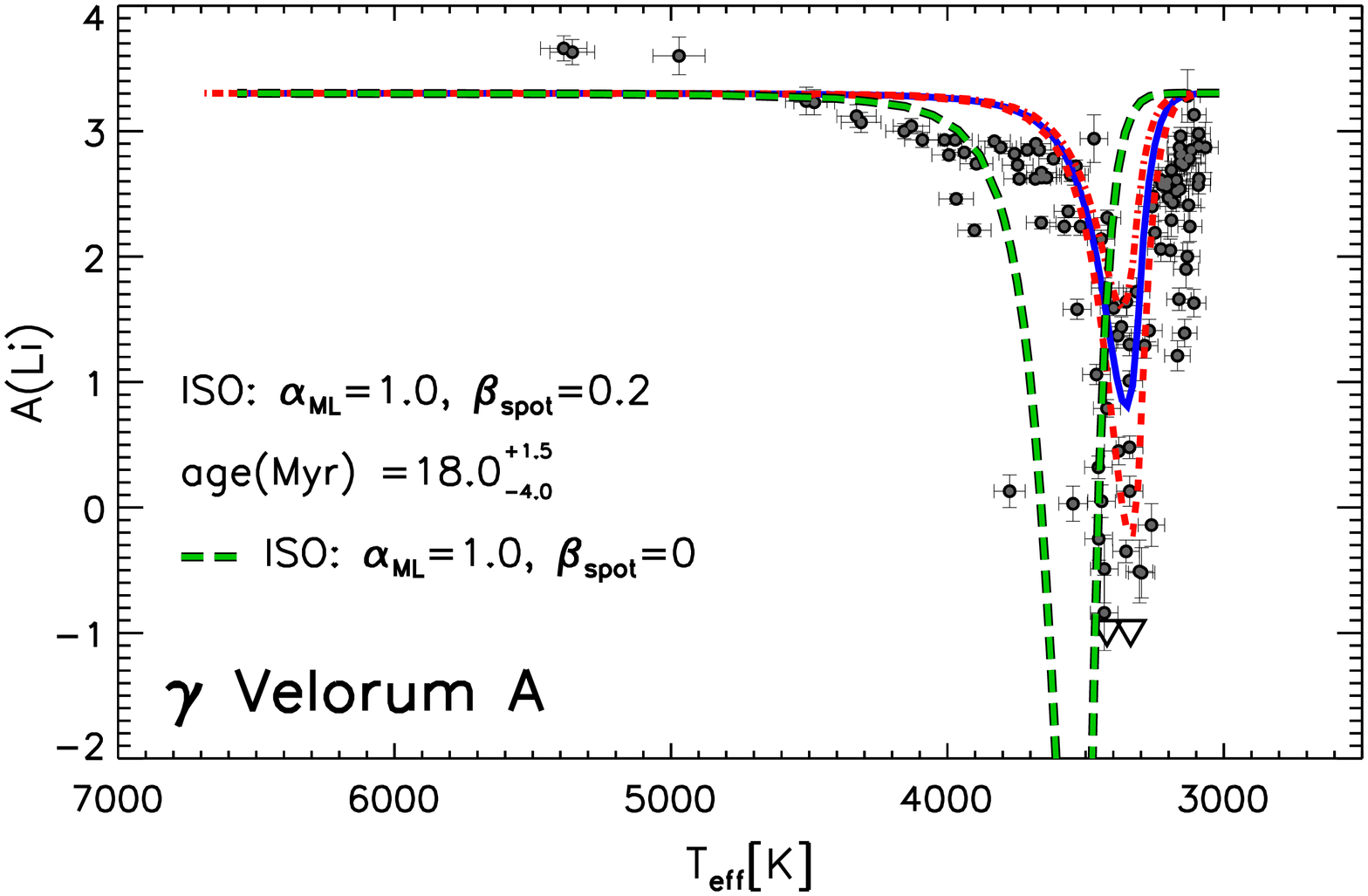}
}
\resizebox{\hsize}{!}{%
\includegraphics{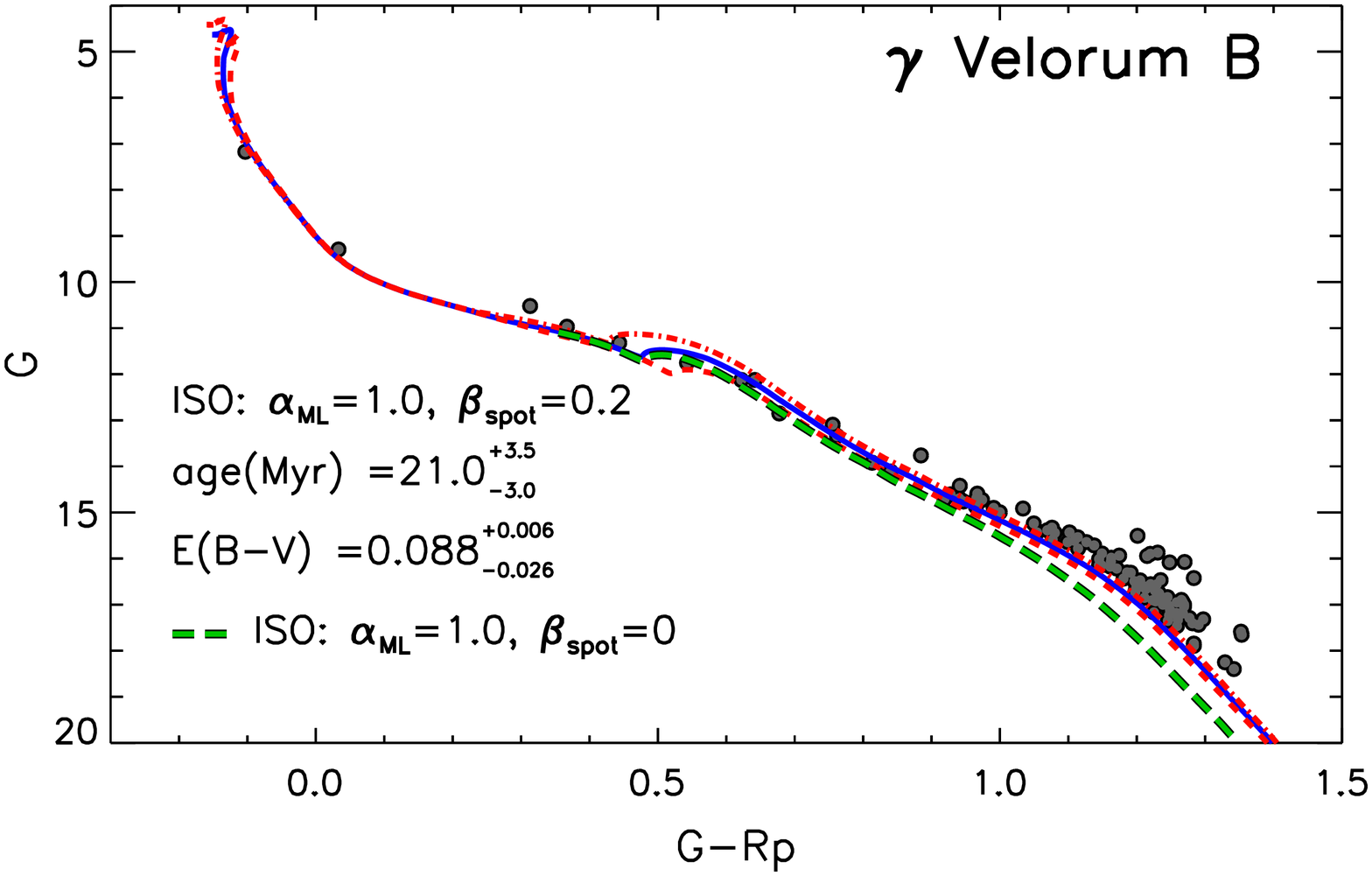}
\includegraphics{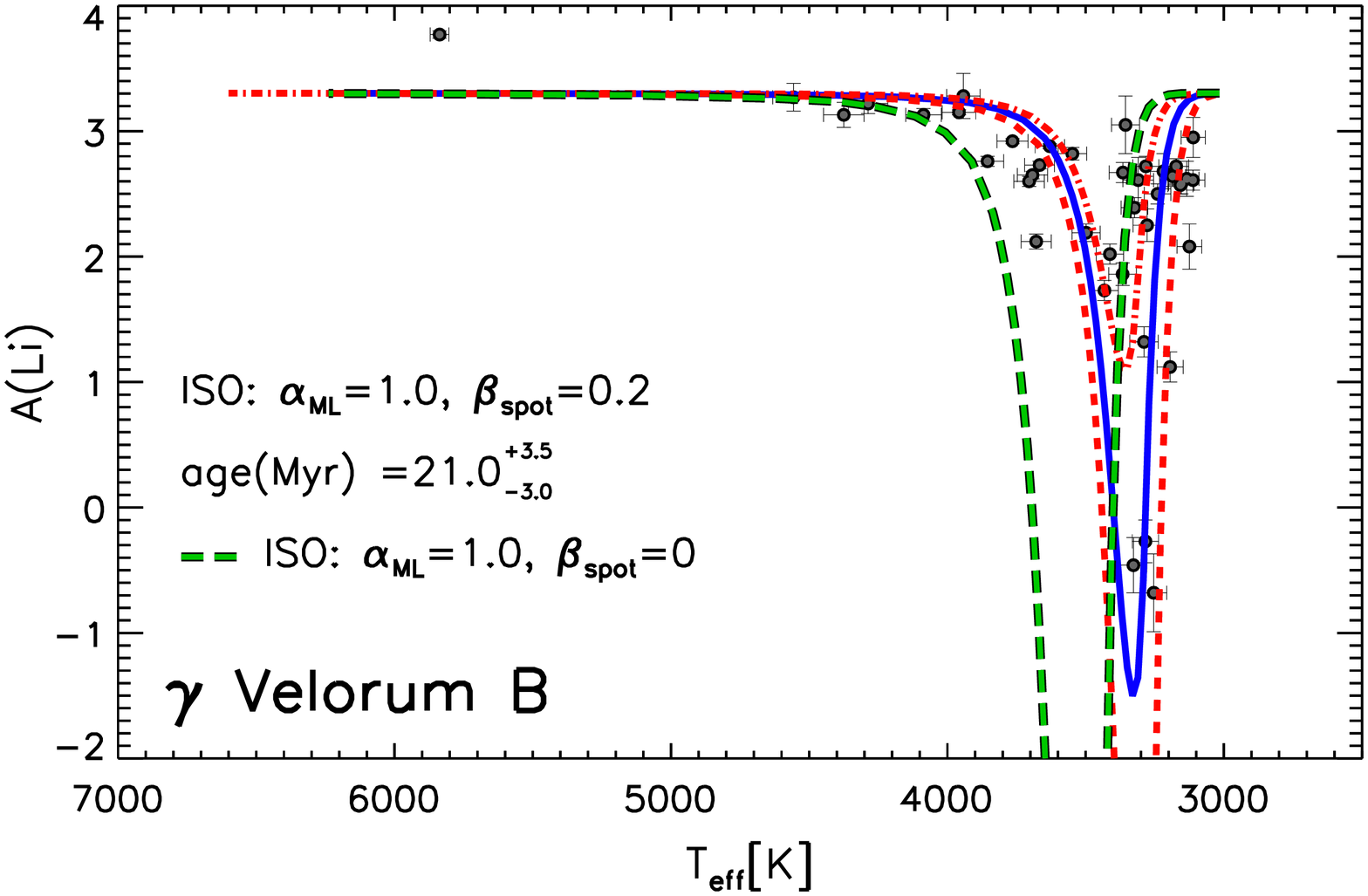}
}
\caption{Comparison of the data with the best-fit model in the CMD ({\it
left panels}) and in the lithium-$T_\mathrm{eff}$ plane ({\it right
panels}) for Gamma~Vel~A ({\it top}) and Gamma~Vel~B ({\it bottom}).
Errors on photometry are smaller than the symbol size. Open downward
triangles in the lithium diagrams represent upper limits for $A$(Li). The
best-fit model with $\alpha_\mathrm{ML}=1.0$ and $\beta_\mathrm{spot}=0.2$
is shown as a solid blue line, and the isochrones corresponding to the
maximum and minimum age derived by the fitting method are shown as dashed
and dot-dashed red lines, respectively. The long-dashed green line is the
isochrone at the same age, reddening, and $\alpha_\mathrm{ML}$ as the
best-fit model, but with $\beta_\mathrm{spot}=0$.}
\label{fig:bf_gvel}
\end{figure*}

The recovery of the age and reddening was performed as follows: we first
derived the age and reddening of each cluster for the reference case. The
reddening was mainly determined by the comparison between theory and 
observation of the zero-age main-sequence (ZAMS) portion of the cluster
sequence because its position in the CMD is independent of age. This region
is populated by stars without a convective envelope that are not affected by
variations in the mixing length or by surface spots. Therefore the 
reddening can be derived independently of the considered model set. For 
this reason, we fixed the reddening to that derived using the reference set
to save computational time. On the other hand, the estimated age changes
using different sets of models because it is derived from the PMS low-mass
stars, which have convective envelopes, thus their position in the CMD
changes if $\alpha_\mathrm{ML}$ varies and if surface spots are included or
excluded. These clusters do not have enough stars in the turn-off region to
allow an age determination from the hydrogen exhaustion luminosity.

\begin{figure*}[!t]
\centering
\resizebox{\hsize}{!}{%
\includegraphics{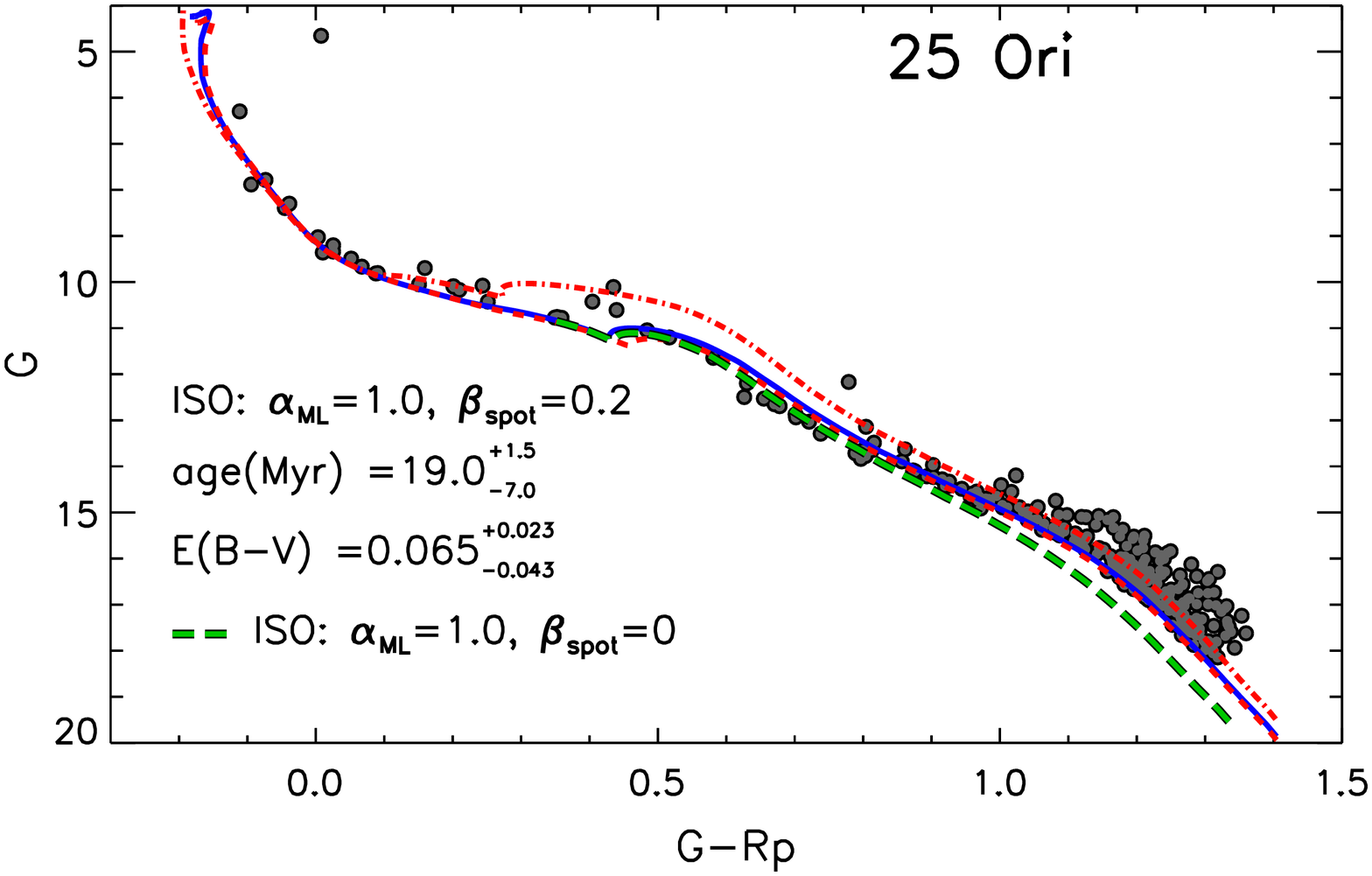}
\includegraphics{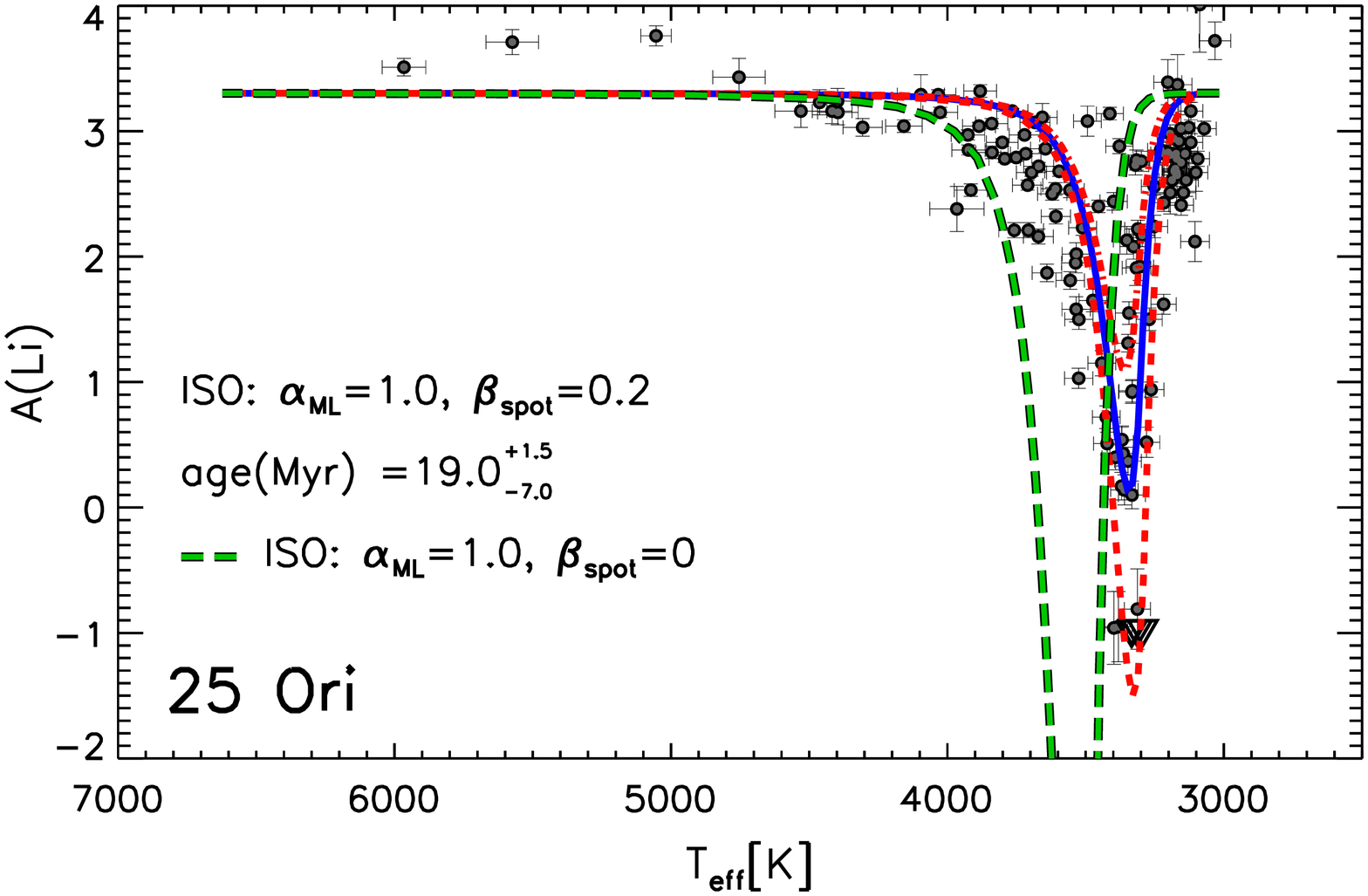}
}
\caption{Same as Fig.~\ref{fig:bf_gvel}, but for the 25~Ori cluster.}
\label{fig:bf_25ori}
\end{figure*}

After the age and reddening for each cluster and each set of models were
derived from the Bayesian fitting of the CMD, we extracted the corresponding
lithium isochrones and compared them with the observed lithium depletion
pattern. This allowed us to identify the model that better reproduces the
data and to consequently determine the best cluster age. The observed
lithium abundances show too much scatter to allow the use of the Bayesian 
maximum likelihood method to derive the characteristics of the best-fit 
model. We also note that if magnetic phenomena are present, they are 
expected to show a distribution of their efficiencies among the cluster 
stars, and this could strongly contribute to the observed dispersion. In 
this case, lithium observations cannot be fitted by a single isochrone, but
the set that best reproduces the average lithium profile can still be
identified. The model parameters given in Table~\ref{tab:model_fit} (and the
corresponding isochrones plotted in
Figs.~\ref{fig:bf_gvel}-\ref{fig:bf_n2516}) reproduce the CMD and the
position of the lithium chasm best. The observed spread in the observed
lithium abundances inside the chasm is too large in most cases to be
reproduced by a single isochrone with a defined set of parameters.

\subsection{Gamma~Vel~A and B}
\label{sec:gvelAB}

The best-fit models for Gamma~Vel~A and Gamma~Vel~B are shown in
Fig.~\ref{fig:bf_gvel}. In both cases, the best agreement with the
observations is obtained when the effect of the magnetic field both
on the stellar interior, with a reduced convection efficiency
$\alpha_\mathrm{ML}=1.0$, and on the surface, with a 20\% effective spot
coverage, is considered. In particular, $\beta_\mathrm{spot}=0.2$ is
required to reproduce the correct position in $T_\mathrm{eff}$ of the
lithium depletion region. Lower or higher values of $\beta_\mathrm{spot}$ in
fact shift the position of the chasm towards higher or lower temperatures,
respectively, as shown in Fig.~\ref{fig:iso_ml_beta}. Higher values of
$\alpha_\mathrm{ML}$ result in significantly younger best-fit ages and lower
lithium depletion than observed. We note, however, that as mentioned
before, the abundance spread in this region is large and cannot be
reproduced using a single isochrone with a fixed parameter set.

In the case of Gamma~Vel~A, we derive a reddening
$E(B-V)=0.062^{+0.006}_{-0.022}$~mag and an age of $18.0^{+1.5}_{-4.0}$~Myr.
The best-fit model can satisfactorily reproduce the average data
distribution in the CMD and the lithium-$T_\mathrm{eff}$ plane. However, the
lithium diagram shows a significant discrepancy at $T_\mathrm{eff}\sim
3700-4300$~K, where the data show much more lithium depletion than
predicted. This might suggest that magnetic effects in these stars are less
efficient. To verify this hypothesis, we also plot in Fig.~\ref{fig:bf_gvel}
the isochrone with $\alpha_\mathrm{ML}=1.0$ and the same age and reddening
as the best-fit one but without spots. A range of spot coverage between 0
and 0.2 could explain the observed higher depletion level at least below
4000~K, including the depth of the depletion pattern. In particular, the
hottest of the strongly depleted stars appear to coincide with the boundary
for $\beta_\mathrm{spot}=0$. This model also seems to provide a better fit
to part of the stars in the CMD for $G-G_\mathrm{RP}=0.5-0.9$. However, even
the model without spots is not able to completely solve the disagreement
above 4000~K. The reason for this is not clear and deserves further
investigation in future works. In addition to this, also stars around
$\sim$\,3200~K appear to be slightly more depleted than the prediction of
the best-fit model. These stars could be reproduced assuming a higher spot
coverage, up to $\beta_\mathrm{spot}=0.4$ (not shown in
Fig.~\ref{fig:bf_gvel}, but see Fig.~\ref{fig:iso_ml_beta}, bottom left
panel). 

We find similar results for Gamma~Vel~B. For this cluster we derive a
reddening $E(B-V)=0.088^{+0.006}_{-0.026}$~mag and an age of
$21.0^{+3.5}_{-3.0}$~Myr. The central age for Gamma~Vel~B is slightly
older than the age derived for Gamma~Vel~A, but it is consistent within the
uncertainties, therefore the two populations could be coeval. The lower
number of stars available for this cluster is responsible for the higher
uncertainty on the best-fit age. Similarly to the previous case, we find
that the model and the mean data
distribution for the CMD and the lithium pattern agree well. In particular, the
model can reproduce the position in $T_\mathrm{eff}$ of the
lithium depletion region very well. As for Gamma~Vel~A, a mix of stars with
different starspot coverage (from no spots to $\beta_\mathrm{spot}=0.2$)
is able to reproduce the observed spread in lithium between $\sim\,$3500
and $\sim\,$4000~K.

We note that, while for both clusters the agreement in the CMD is very good
for $G-G_\mathrm{RP} \la$\,0.9--1, the best-fit model for cooler stars does
not provide a perfect match to the cluster sequences, being consistent with
their lower envelope. This discrepancy could in part be due to the presence
of stars with higher spot coverages, as suggested by the lithium pattern,
which would make them redder. As mentioned in Sect.~\ref{sec:age}, however,
there might be other potential effects that are not well understood so far,
that might also play a role and might partly contribute to the discrepancy.

Our results agree very well with what was found by \citet{jeffries17}, who
showed that models with inflated radii from magnetic inhibition of
convection or starspots are required to simultaneously reproduce the CMD and
lithium pattern of Gamma~Vel~A and B at an age of 18$-$21~Myr. However,
\citet{jeffries17} did not distinguish between the two populations. To our
knowledge, ours is the first precise determination of the ages and reddening
of Gamma~Vel~A and B separately.

\subsection{25~Ori}
\label{sec:25ori}
Figure~\ref{fig:bf_25ori} shows the best-fit isochrone for the 25~Ori
cluster. As in the case of Gamma~Vel~A and B, the best agreement in
the CMD and the lithium-$T_\mathrm{eff}$ plane is obtained using the
models with $\alpha_\mathrm{ML}=1.0$ and $\beta_\mathrm{spot} = 0.2$. We
derive an age of $19.0^{+1.5}_{-7.0}$~Myr and a reddening
$E(B-V)=0.065^{+0.023}_{-0.043}$. This places this cluster in the same age
range as Gamma~Vel A and B, as expected from the very similar lithium
pattern. As already found for Gamma~Vel, the age derived using magnetic
models is much higher than the values derived previously from simple
isochrone fitting with standard models, which range from $6.1\pm 0.8$~Myr
\citep{downes14} to $13.0\pm 1.3$~Myr \citep{kos19}. The best fit with our
standard model would give an age of 12~Myr, consistent with previous
determinations, but this model would not be able to reproduce the observed
lithium pattern.

As already observed in the previous section, the best-fit model alone is not
able to completely reproduce the lower-mass part of the CMD and the observed
abundance spread in this case either. In particular, several stars around
$3600-4000$~K are more depleted than predicted by the best-fit isochrone.
These stars could be reproduced assuming a range of spot coverages between 0
and 0.2, while higher spot coverage fractions could better explain the
abundances of the coolest stars, as in Gamma~Vel~A.

\begin{figure*}[!ht]
\centering
\resizebox{\hsize}{!}{%
\includegraphics{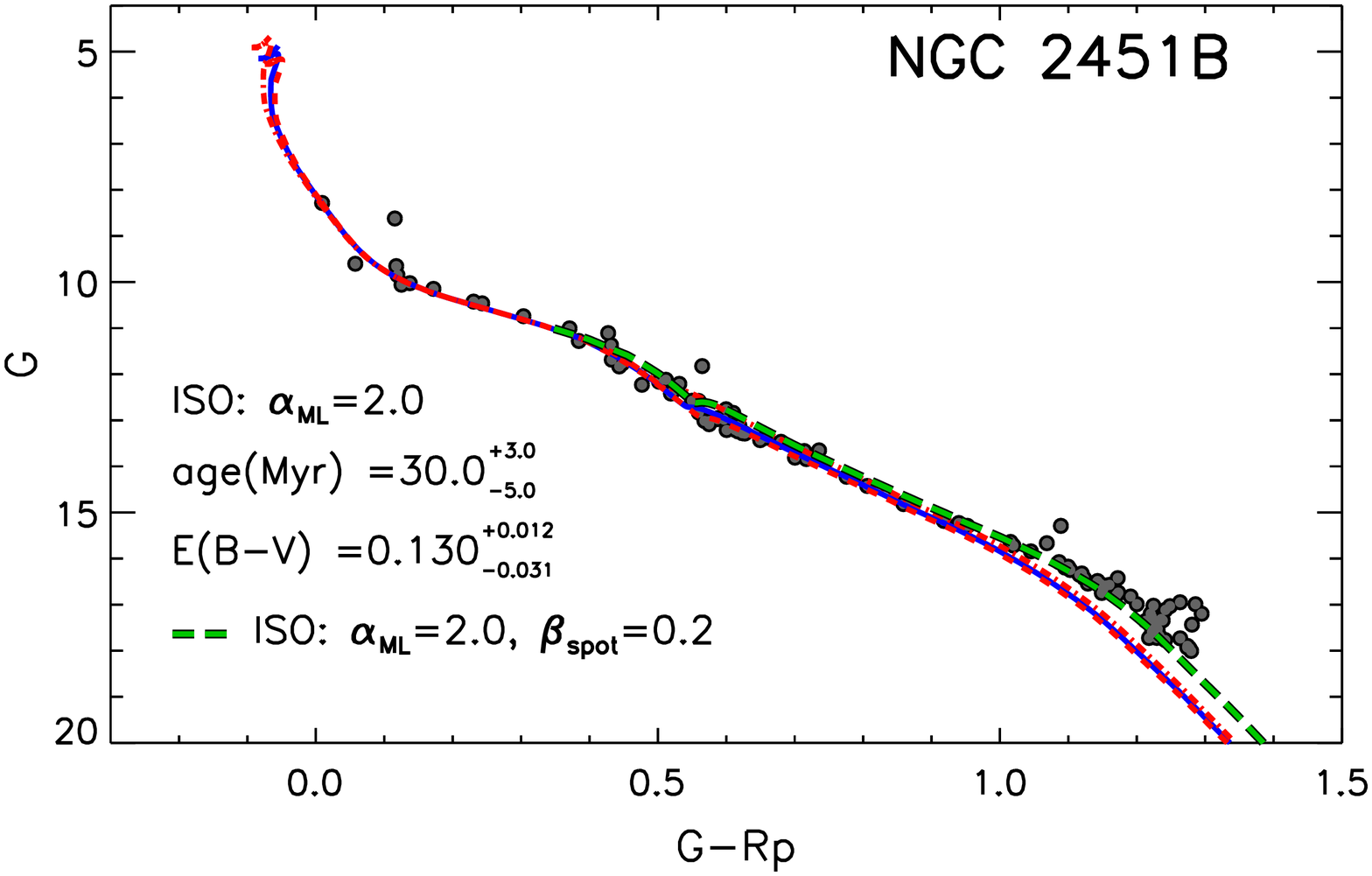}
\includegraphics{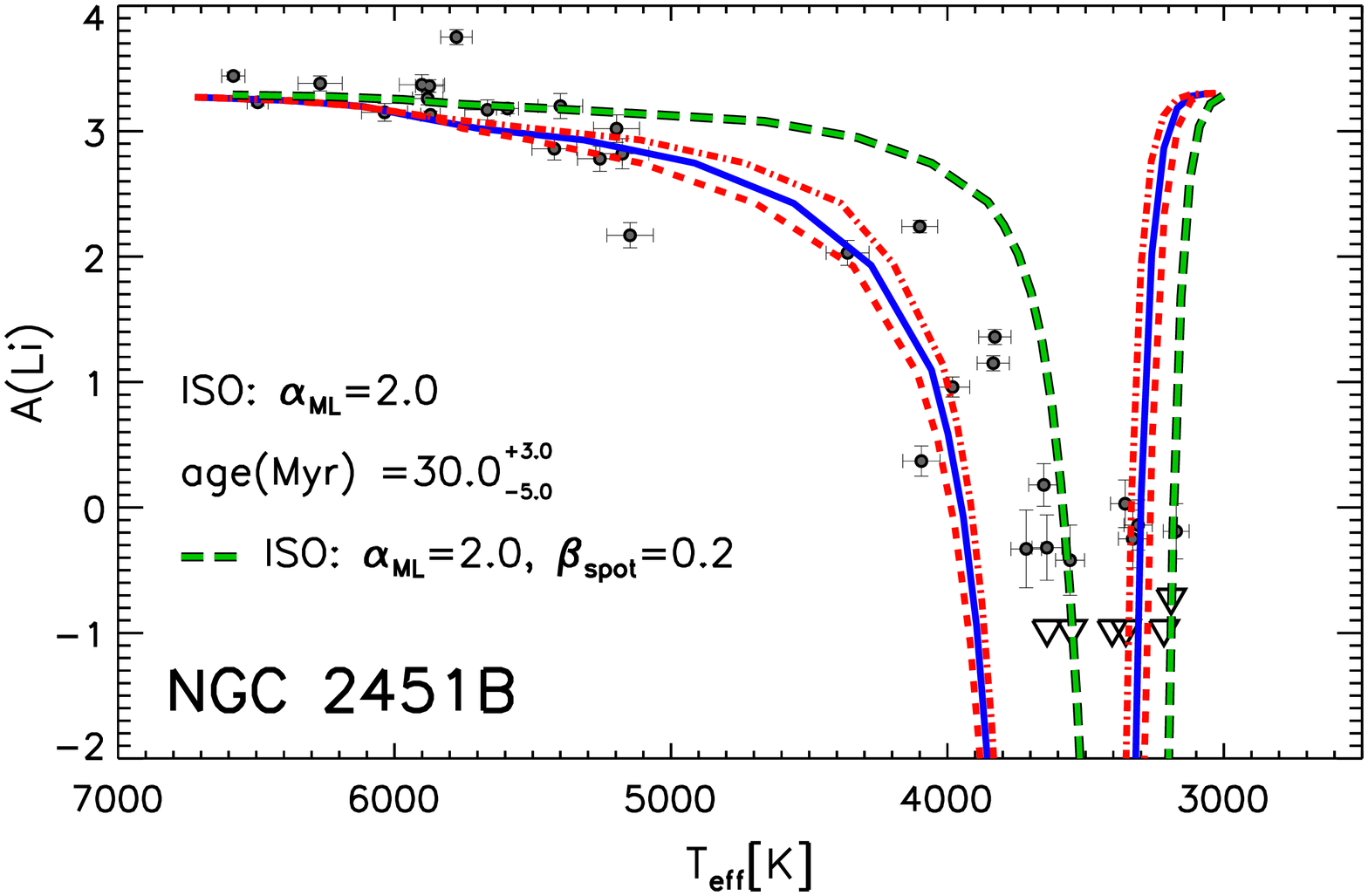}
}
\resizebox{\hsize}{!}{%
\includegraphics{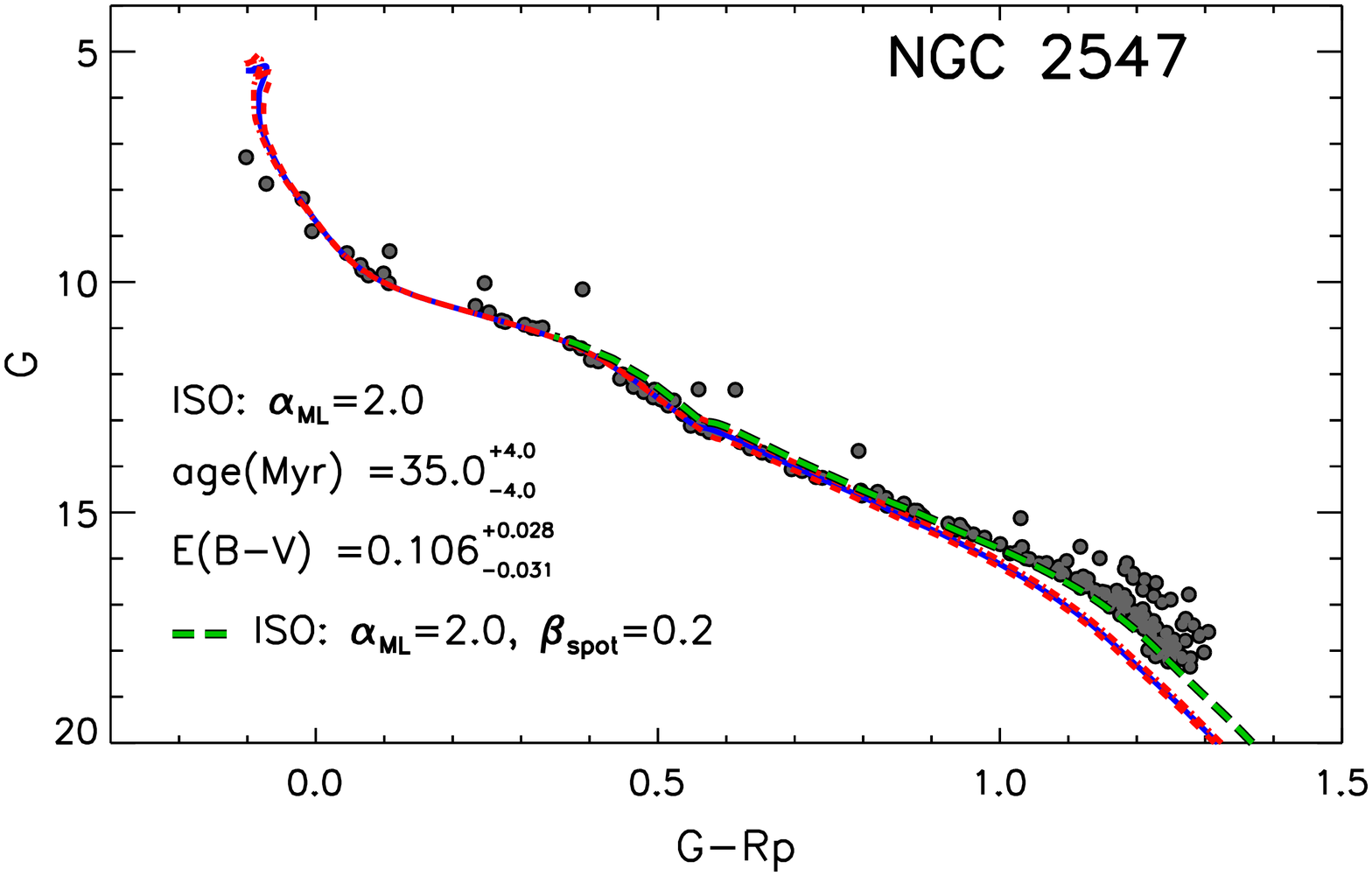}
\includegraphics{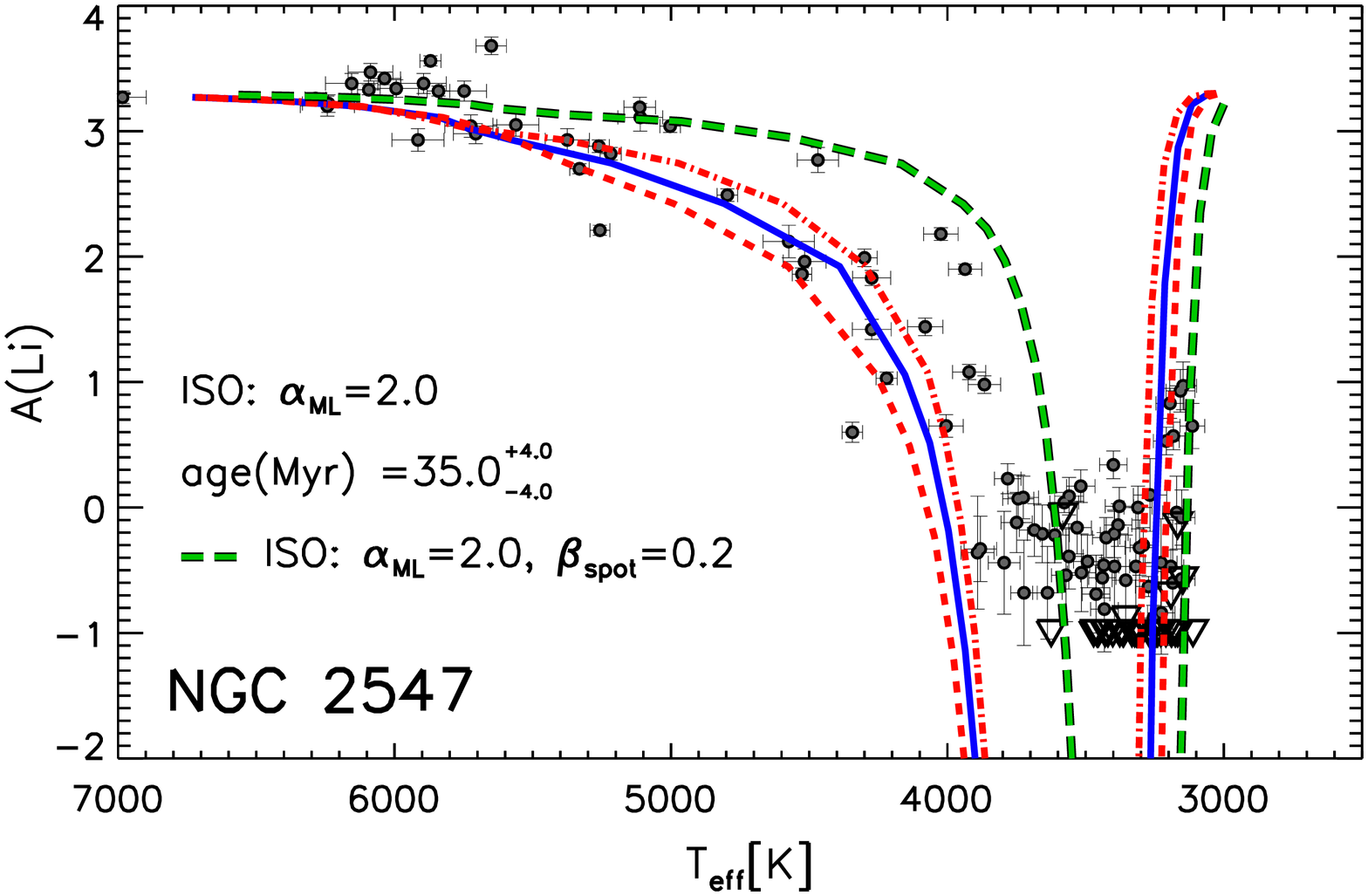}
}
\caption{Same as Fig.~\ref{fig:bf_gvel} for NGC~2451\,B ({\it top}) and
NGC~2547 ({\it bottom}). In this case, the best-fit model (solid blue line)
corresponds to $\alpha_\mathrm{ML}=2.0$ and $\beta_\mathrm{spot}=0$. The
long-dashed green line shows the isochrone with the same age, reddening, and
$\alpha_\mathrm{ML}$ as the best-fit model, but computed with
$\beta_\mathrm{spot}=0.2$.}
\label{fig:bf_n2451b_2547}
\end{figure*}

\subsection{NGC~2451\,B and NGC~2547}
\label{sec:ngc2451_2547}

Figure~\ref{fig:bf_n2451b_2547} shows the best-fit results for the two
nearly coeval clusters NGC~2451\,B and NGC~2547. In both cases, the model
that provides the best agreement with the data in the CMD and the
lithium-$T_\mathrm{eff}$ plane is the model with solar-calibrated mixing
length, that is, $\alpha_\mathrm{ML}=2.0$, without spots. Models with reduced
convection efficiency are not able to reproduce the lithium pattern at
$T_\mathrm{eff}\ga 4000$~K because they provide too little lithium
depletion.

In the case of NGC~2451\,B, we derive a reddening $E(B-V) =
0.130^{+0.012}_{-0.031}$~mag and an age of $30.0^{+3.0}_{-5.0}$~Myr, in
agreement with the recent results by \citet{randich18}. This cluster has a
limited number of members as well, which prevents a precise age
determination. Nevertheless, the agreement with the observations is very
good throughout the entire CMD sequence, except for the very low-mass tail. 
For NGC~2547 we derive a reddening $E(B-V)=0.106^{+0.028}_{-0.031}$~mag
and an age of $35.0^{+4.0}_{-4.0}$~Myr, consistent with the LDB age
derived by \citet{jeffries05} and with other previous studies
\citep[e.g.][]{naylor06,randich18}. Similarly to NGC~2451\,B, standard
isochrones provide the best fit to the CMD, except at very low masses. The
standard model also reproduces the lower envelope of the lithium depletion
pattern in both clusters well, as well as the position of the LDB in
NGC~2547. As mentioned in Sect.~\ref{sec:finalmem}, however, both clusters
show significant scatter, and several stars are less depleted than expected.
This scatter, as well as the low-mass tail in the CMD, could be accounted
for if these stars were still affected by a non-negligible magnetic effect,
which in such low-mass stars is reflected only in a certain percentage of
spot coverage. Figure~\ref{fig:bf_n2451b_2547} indeed shows that the
isochrone with the same age, reddening, and $\alpha_\mathrm{ML}$ as the
best-fit isochrone, but with $\beta_\mathrm{spot}=0.2$, provides a better
match to the lower cluster sequences and can successfully reproduce the
upper envelope of the lithium pattern for both clusters. The spotted model
can also account for the small spread observed among the stars at the LDB in
NGC~2547.

In NGC~2547 some stars at lower temperatures inside the chasm still retain
some lithium, in contrast to what is expected from the models. It is
possible that these stars are more magnetically active and have a larger
spot coverage, which would lead to a lower depletion efficiency. We note,
however, that lithium measurements of strongly depleted stars at these low
temperatures are difficult because of the significant blending with
molecular bands. Therefore we cannot exclude that the observed scatter might
at least in part be due to uncertainties in the measures.

\begin{figure*}
\centering
\resizebox{\hsize}{!}{%
\includegraphics{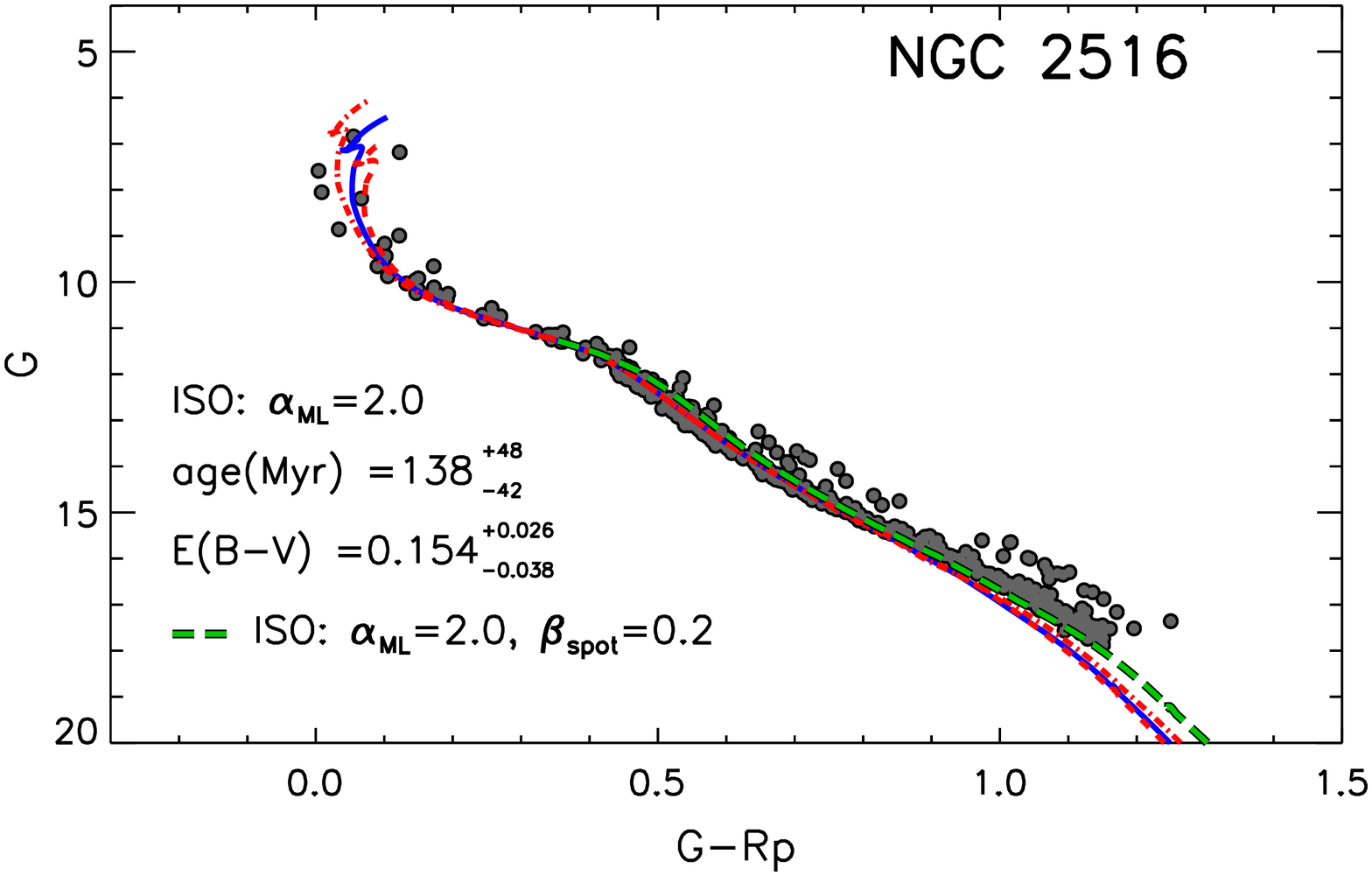}
\includegraphics{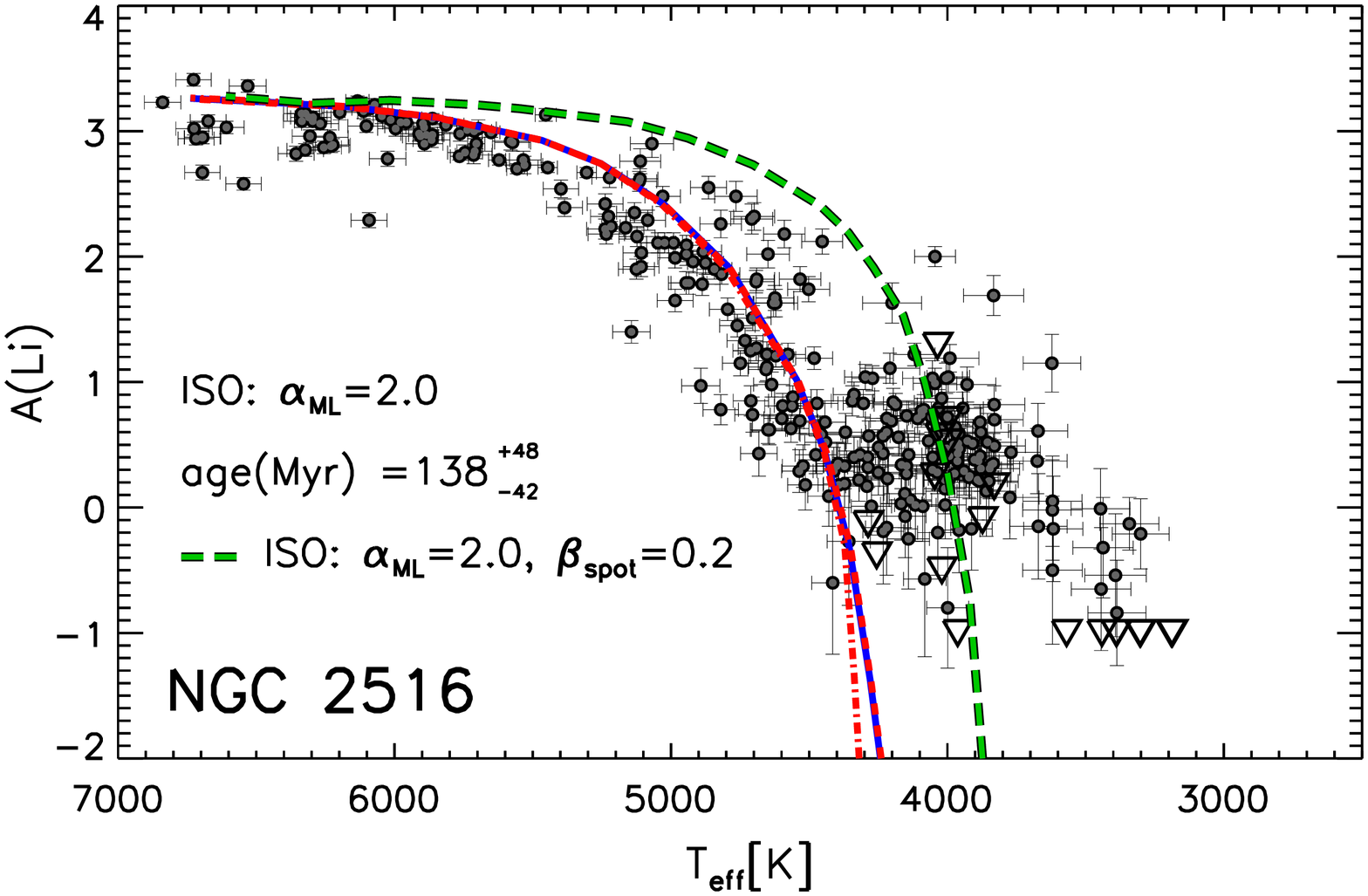}
}
\caption{Same as Fig.~\ref{fig:bf_n2451b_2547}, but for the NGC~2516
cluster.}
\label{fig:bf_n2516}
\end{figure*}

\subsection{NGC~2516}
\label{sec:ngc2516}

NGC~2516 is the oldest cluster in our sample. We derive an age of
$138^{+48}_{-42}$~Myr and a reddening $E(B-V)=0.154^{+0.026}_{-0.038}$~mag
using the solar-calibrated isochrones (Fig.~\ref{fig:bf_n2516}). The age and
reddening we found are consistent with the values given in the literature
\citep[see e.g.][and references
therein]{jeffries98,sung02,lyra06,randich18}. With the exception of the
low-mass tail, the CMD is reproduced very well. The age in this cluster is
derived mainly from turn-off stars, as low-mass stars are on the ZAMS and
consequently, their position is independent of age. Unfortunately, 
as noted in Sect.~\ref{sec:finalmem}, bright stars show a significant
photometric scatter, so that the derived age uncertainty is quite large. 
The standard model isochrone also agrees well with the bulk of
the lithium distribution for $T_\mathrm{eff} \ga 4500$~K, although
significant scatter is present below $\sim$\,5000~K. This scatter can be
accounted for by assuming a distribution of the spot coverage among the
cluster stars between 0 and 20\%. In particular, the model with
$\beta_\mathrm{spot}=0.2$ provides a good match to the upper envelope of the
lithium pattern above $\sim\,$4000~K. This model also appears to better
reproduce the low-mass tail of the CMD, similarly to NGC~2451\,B and
NGC~2547. In this case as well, a number of cooler stars has much higher
lithium abundances than predicted by the models. This might be
reproduced by assuming a higher spot coverage fraction.

\section{Discussion and conclusions}
\label{sec:disc}

Observations of field and cluster stars suggest that effects due to magnetic
activity, which can reduce the external convection efficiency and produce a
significant coverage of the stellar surface with spots, might be present.
In this paper, we modelled these effects in a consistent way including them
in the Pisa stellar evolutionary code. The code was used to calculate
cluster isochrones and to predict the time behaviour of the surface lithium
abundance. Theoretical predictions were compared with the observed CMDs and
lithium abundance patterns of five young open clusters of different ages,
taking advantage of the homogeneous stellar parameters and lithium
abundances provided by GES, and of the precise astrometry and photometry
from the {\em Gaia} EDR3 release. The quality of the combined GES and {\em
Gaia} data allowed us to gain important insights into PMS evolution.

Lithium data place strong constraints on the best set of models \citep[see
also][]{jeffries21}. As shown in the previous section, lithium depletion
strongly depends on a change in $\alpha_\mathrm{ML}$ and
$\beta_\mathrm{spot}$ and, more importantly, the effects of these
parameters are different. Intermediate-mass stars are affected by a
variation in the convection efficiency, but are insensitive to spots. On
the other hand, in low-mass stars, which are almost adiabatic, lithium
abundances are not affected by a variation in $\alpha_\mathrm{ML}$, but
strongly depend on the adopted effective spot coverage. This differential
sensitivity allows us to distinguish among models that were computed with
different assumptions on these parameters.

From the analysis of the lithium abundance patterns and the CMDs of our
sample of young open clusters, the following picture seems to emerge: 
\begin{itemize}
\item For clusters younger than $\sim\,$20~Myr, we confirm that standard
models cannot simultaneously reproduce the observed CMD and the lithium
pattern, as already noted by \citet{jeffries17} for Gamma~Vel. These
clusters can be reproduced using models that include both a non-negligible
spot coverage and a reduced super-adiabatic convection efficiency in the
stellar envelope. In particular, a mix of different spot coverages in
different stars that increase with decreasing mass appears to be necessary
to explain the observed lithium depletion pattern.
\item In older clusters (i.e. ages older than $\sim$\,30$-$40~Myr),
standard isochrones provide a good agreement with the data. However,
low-mass stars in these clusters show a significant discrepancy with the
models in both the CMD and the lithium pattern. This discrepancy can be
reduced when models with spots are considered. Spotted models can not only
explain the observed lithium dispersion in NGC~2451\,B, NGC~2547, and
NGC~2516, in particular the large spread at
$T_\mathrm{eff}\sim$\,3200$-$3800~K, but they also provide a better fit to
the low-mass tail in the CMD. The value of $\beta_\mathrm{spot}=0.2$ that
best reproduces the low-mass cluster sequences and the upper envelope of the
lithium patterns is consistent with the 34\% real spot coverage found by
\citet{jeffries21} to reproduce the lithium upper envelope in M35, assuming
$T_\mathrm{spot}/T_\mathrm{eff}=0.8$. 
\end{itemize}

\begin{figure}
\centering
\resizebox{\hsize}{!}{\includegraphics{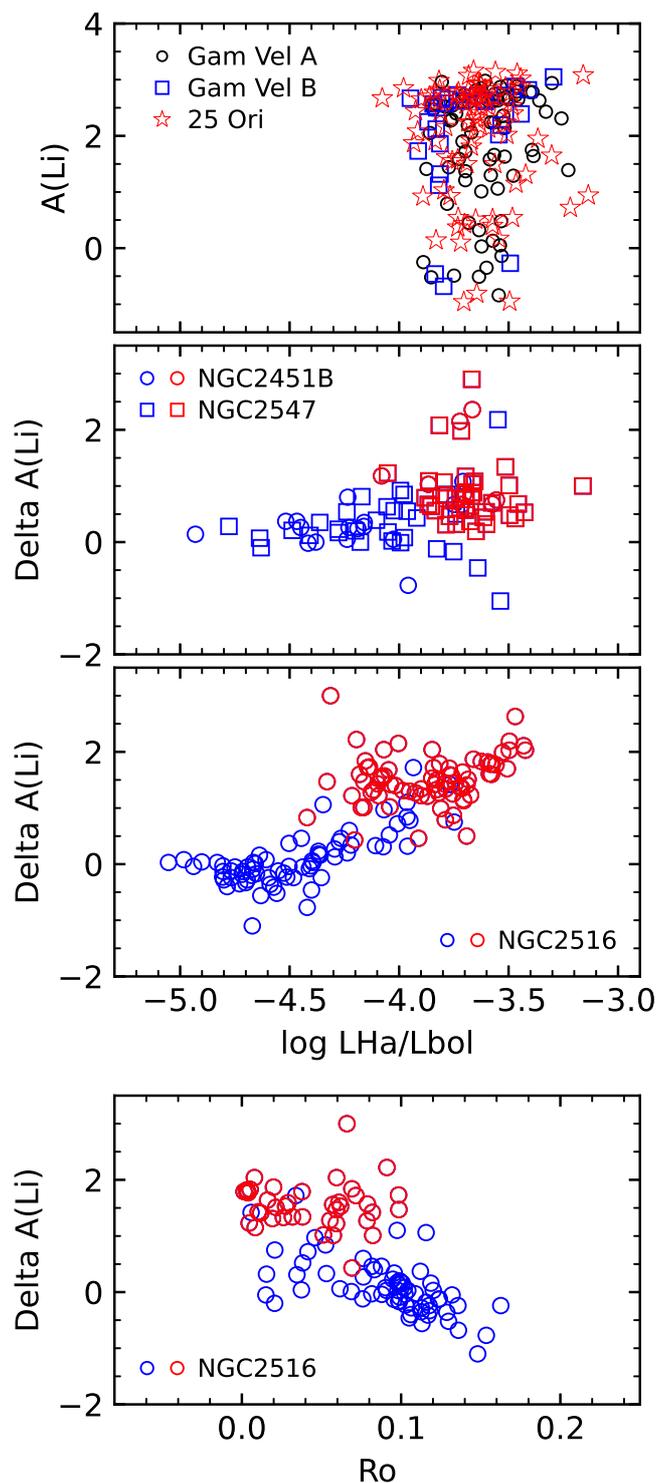}} 
\caption{Lithium as a function of the chromospheric activity
indicator $\log L_\mathrm{H\alpha}/L_\mathrm{bol}$ for detections only (top
three panels). In the first panel, we plot $A$(Li) for stars with
3000\,$<T_\mathrm{eff}<$\,3700~K in Gamma~Vel~A (open black circles),
Gamma~Vel~B (open blue squares) and 25~Ori (red stars). In the second and
third panels, we show the lithium excess $\Delta A$(Li) (see text) for
stars with 3000\,$<T_\mathrm{eff}<$\,6000~K in NGC~2451\,B and NGC~2547
(second panel, open circles and squares, respectively), and NGC~2516
(third panel, open circles). The red symbols mark stars for which
the models predict full lithium depletion. 
{\it Bottom panel:} $\Delta A$(Li) for NGC~2516 members as a
function of the Rossby number $R_\circ$ from \citet{fritzewski20}. The
symbols are the same as in the third panel.}
\label{fig:correl}
\end{figure}

Magnetically active stars are well known to exhibit spots covering a
significant fraction of their surface. Recently, \citet{fang16}
investigated the spot distribution of low-mass stars in the Pleiades,
finding a large spread, from no spots to high coverage fractions: the
upper envelope of this distribution increases from G- to M-type stars, up
to maximum values of effective coverage $\sim$\,0.3 for stars in the
saturated regime. However, the spread in spot coverage is present even in
the saturated regime for stars with $T_\mathrm{eff}\le 3800$~K. These
results strongly support our hypothesis that the lithium dispersion might be
due to the presence of stars with different spot coverage fractions that
increase at lower masses. 

As a further verification, we investigated whether the lithium abundances are
correlated with magnetic activity. It is well known that magnetic activity
increases with increasing rotation rate (or equivalently, decreasing Rossby
number $R_\circ$) up to a saturation level corresponding to $R_\circ\la
0.1$ \citep[e.g.][]{pizzol03}. A correlation between lithium and rotation
has long been known to exist in the Pleiades \citep{soderblom93,somers17}
and has also been observed  in other clusters even at very young ages
\citep[e.g.][]{bouvier16,jeffries21}. In Fig.~\ref{fig:correl} we show the
lithium abundances as a function of the chromospheric H$\alpha$ emission
$\log L(\mathrm{H}\alpha)/L_\mathrm{bol}$, which was computed from the
measured chromospheric H$\alpha$ flux available in the GES catalogue. In
the case of NGC~2516, we also show lithium as a function of the Rossby
number provided by \citet{fritzewski20}. For the three younger clusters we
simply plot the measured $A$(Li) for stars between 3000 and 3700~K. For
the older clusters, we instead computed the difference $\Delta A$(Li) between
the observed abundance and that predicted by the best-fit standard model
for stars between 3000 and 6000~K (excluding those close to the LDB),
assuming the value $A\mathrm{(Li)}=-1$ for complete depletion, which is
consistent with the minimum value measured in GES. Stars that should be
fully depleted according to the model are marked with red symbols in
Fig.~\ref{fig:correl}. The lithium excesses for fully depleted stars should
be regarded as lower limits because their true abundance could be lower than
our assumed minimum value. We find a clear correlation with chromospheric
activity and Rossby number for NGC~2516: the lithium excess increases with
decreasing $R_\circ$ and increasing $\log
L(\mathrm{H}\alpha)/L_\mathrm{bol}$, and stars with the highest excess,
including the undepleted M-type stars, all have $R_\circ \la 0.1$, that is,
they are saturated. In the younger clusters, all
the M-type stars have saturated levels of H$\alpha$ emission, implying
that they have strong magnetic activity.

In conclusion, models including the effects of magnetic activity, such as
starspots, seem able to provide a consistent explanation of the low-mass
cluster sequences and of the lithium abundance patterns observed in young
clusters. The observed dispersion in the lithium abundances could be the
result of different star-to-star spot coverage fractions. However, some
inconsistencies are still present, both at the lower-end of the CMD and
between the predicted and observed lithium pattern. A possible improvement
could be attained by introducing a mass-dependent spot coverage, or a
variation in the spot coverage in time due to the evolution of rotation in
low-mass stars. We cannot exclude either that some other mechanism may play a
role in producing part of the observed lithium dispersion. This requires
further investigation and a comparison with a larger sample of young open
clusters.

\begin{acknowledgements}
We thank E. Martin for his fruitful comments as referee of this paper.
Based on data products from observations made with ESO Telescopes at the La
Silla Paranal Observatory under programmes 188.B-3002, 193.B-0936, and
197.B-1074. These data products have been processed by the Cambridge
Astronomy Survey Unit (CASU) at the Institute of Astronomy, University of
Cambridge, and by the FLAMES/UVES reduction team at INAF/Osservatorio
Astrofisico di Arcetri. These data have been obtained from the Gaia-ESO
Survey Data Archive, prepared and hosted by the Wide Field Astronomy Unit,
Institute for Astronomy, University of Edinburgh, which is funded by the UK
Science and Technology Facilities Council. This work was partly supported by
the European Union FP7 programme through ERC grant number 320360 and by the
Leverhulme Trust through grant RPG-2012-541. We acknowledge the support from
INAF and Ministero dell'Istruzione, dell'Universit\`a e della Ricerca (MIUR)
in the form of the grant "Premiale VLT 2012". The results presented here
benefit from discussions held during the Gaia-ESO workshops and conferences
supported by the ESF (European Science Foundation) through the GREAT
Research Network Programme.
We acknowledge the support from INAF in the form of the grant for mainstream
projects ``Enhancing the legacy of the Gaia-ESO Survey for open clusters
science''.
E.T. is funded by the Czech Science Foundation GA\u{C}R (Project:
21-16583M). He also thanks the University of Pisa for the hospitality during
6/09/2021--5/10/2021 through the "visiting fellow program". 
P.G.P.M. and S.D. acknowledge INFN (Iniziativa specifica TAsP). 
E.P. and N.S. acknowledge financial support from Progetto Main Stream INAF
``Chemo-dynamics of globular clusters: the Gaia revolution''.
R.B. acknowledges financial support from the project PRIN-INAF 2019 
``Spectroscopically Tracing the Disk Dispersal Evolution''.
F.J.E. acknowledges financial support from the Spanish MINECO/FEDER through
the grant AYA2017-84089 and MDM-2017-0737 at Centro de Astrobiología
(CSIC-INTA), Unidad de Excelencia María de Maeztu, and from the European
Union’s Horizon 2020 research and innovation programme under Grant Agreement
no. 824064 through the ESCAPE - The European Science Cluster of Astronomy \&
Particle Physics ESFRI Research Infrastructures project.
This work has made use of data from the European Space Agency (ESA) mission
{\em Gaia} (\url{https://www.cosmos.esa.int/gaia}), processed by the {\it
Gaia} Data Processing and Analysis Consortium (DPAC,
\url{https://www.cosmos.esa.int/web/gaia/dpac/consortium}). Funding for the
DPAC has been provided by national institutions, in particular the
institutions participating in the {\em Gaia} Multilateral Agreement.
This research made use of {\sc Astropy} (http://www.astropy.org), a
community-developed core Python package for Astronomy
\citep{astropy13,astropy18}.
\end{acknowledgements}

\bibliographystyle{aa} 
\bibliography{biblio}

\begin{appendix}

\section{Maximum likelihood fit of the distribution of RVs, parallaxes,
and proper motions}
\label{sec:mlfit}

\begin{figure}
\centering
\resizebox{\hsize}{!}{\includegraphics{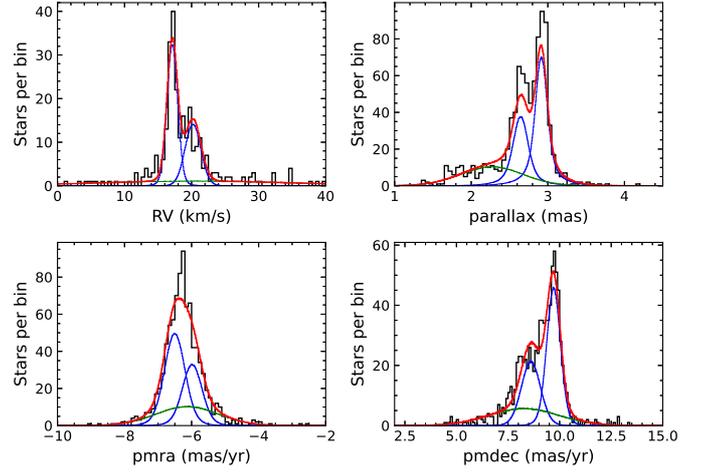}}
\caption{Best-fit distribution (red lines) of Gamma~Vel RVs, parallaxes and
proper motions (black histograms), projected onto the four parameters
planes. The blue lines show the two cluster components, and the green
lines represent the field contribution.}
\label{fig:gvel_mlfit}
\end{figure}

As mentioned in Sect.~\ref{sec:rvastr}, we performed a maximum likelihood
fit%
\footnote{For the fit, we adopted the {\sc slsqp} minimisation method as
implemented in \texttt{scipy}.}
of the distributions of RVs, parallaxes, and proper motions for the full
datasets. We assumed that the total probability distribution is described by
the sum of one or two cluster populations plus a field population. In
particular, we extended the approach described by \citet{lindegren00} that was
applied to {\em Gaia}~DR2 data by \citet{francio18} and \citet{roccat18},
which takes the full covariance matrix and the intrinsic dispersions of each
component into account. Each population $j$ is described by a
four-dimensional multivariate Gaussian that for each star $i$ is given by
\begin{equation}
\mathcal{L}_{ji} = \frac{1}{(2\pi)^2 |\Sigma_{ji}|^{1/2}} \,\exp 
\left[-\frac{1}{2}
(\vec{x_i}-\vec{x_{\circ j}})^T \Sigma_{ji}^{-1} (\vec{x_i}-\vec{x_{\circ j}}) 
\right] \> ,
\end{equation}
where
\begin{equation}
\vec{x_i}-\vec{x_{\circ j}} = \left( 
\begin{array}{c}
v_{r i} - v_{r\circ j} \\
\varpi_i - \varpi_{\circ j} \\
\mu_{\alpha* i}-\mu_{\alpha*\circ j} \\
\mu_{\delta i}-\mu_{\delta\circ j} \\
\end{array}
\right)
\end{equation}
and $\Sigma_{ji} = C_i +\Sigma_{\circ j}$ is the sum of the individual
covariance matrix, 
\begin{equation}
C_i = \left(
\small
\begin{array}{cccc}
\sigma_{{v_r} i}^2 & 0& 0& 0\\
0& \sigma_{\varpi i}^2 & 
\rho_{\varpi\mu_{\alpha*} i}\,\sigma_{\varpi i}\,\sigma_{\mu_{\alpha*} i} &
\rho_{\varpi\mu_\delta i}\,\sigma_{\varpi i}\,\sigma_{\mu_\delta i} \\
0& \rho_{\varpi\mu_{\alpha*} i}\,\sigma_{\varpi i}\,\sigma_{\mu_{\alpha*} i} &
\sigma_{\mu_{\alpha*} i}^2 &
\rho_{\mu_{\alpha*}\mu_\delta i}\,\sigma_{\mu_{\alpha*} i}
\,\sigma_{\mu_\delta i} \\
0& \rho_{\varpi\mu_\delta i}\,\sigma_{\varpi i}\,\sigma_{\mu_\delta i} &
\rho_{\mu_{\alpha*}\mu_\delta i}\,\sigma_{\mu_{\alpha*} i}
\,\sigma_{\mu_\delta i}  &
\sigma_{\mu_\delta i}^2\\
\end{array}
\!\!\right)
\end{equation}
(where $\rho_{\varpi\mu_{\alpha*} i}$, $\rho_{\varpi\mu_\delta i}$,
$\rho_{\mu_{\alpha*}\mu_\delta i}$ are the correlation coefficients
available in the {\em Gaia} archive), and of the matrix of intrinsic
dispersions, 
\begin{equation}
\Sigma_{\circ j} = \left( 
\small
\begin{array}{cccc}
\sigma_{{v_r}\circ j}^2 & 0& 0& 0\\
0& \sigma_{{\varpi}\circ j}^2 & 0& 0\\
0& 0& \sigma_{{\mu_{\alpha*}}\circ j}^2 & 0\\
0& 0& 0& \sigma_{{\mu_\delta}\circ j}^2\\
\end{array}
\right) \>.
\end{equation}
This distribution reduces to three dimensions (astrometry part only) or
one dimension (RV part only) when the RV or the astrometry are missing or
flagged, respectively. The total likelihood for each star is then 
\begin{equation}
\mathcal{L}_i = \sum_{j=1}^{N} f_j \>\mathcal{L}_{ji} \>,
\end{equation}
where $N$ is the number of populations, and $f_j$ is the fraction of stars
belonging to population $j$, with $\sum_{j=1}^{N} f_j =1$.
An example of the best-fit distribution, projected onto the four parameters
planes, is shown in Fig.~\ref{fig:gvel_mlfit} for the case of Gamma~Vel.

\section{Properties of selected members}

Tables~\ref{tab:25ori}-B.6 (available at the CDS) include the properties of
the high-probability members selected for our analysis. Here we show a
portion of Table~\ref{tab:25ori} for 25~Ori as an example.

\begin{table*}
\caption{First and last five lines of the list of high-probability members of
25~Ori}
\label{tab:25ori}
\centering
\small
\begin{tabular}{llccccc}
\hline\hline
\noalign{\smallskip}
GES CNAME & {\it Gaia} EDR3 ID & RA   & DEC  & RV           & $T_\mathrm{eff}$& $A$(Li) \\
          &                    & (deg)& (deg)& (km s$^{-1}$)& (K)             & (dex)   \\
\noalign{\smallskip}
\hline
\noalign{\smallskip}
05225186+0145132& 3234193283936882432& 80.716105& 1.753682& $21.09\pm 0.56$& $3203\pm 50$& $3.39\pm 0.18$\\
05225609+0136252& 3222180741446540544& 80.733684& 1.607000& $19.94\pm 0.35$& $3333\pm 49$& $<-1.00$      \\
05225889+0145437& 3234194039851131008& 80.745381& 1.762147& $20.82\pm 0.27$& $3320\pm 48$& $2.73\pm 0.08$\\
05230387+0134335& 3222180402144875648& 80.766089& 1.576025& $21.24\pm 0.24$& $4158\pm 66$& $3.04\pm 0.05$\\
05230596+0138511& 3222182051412540800& 80.774851& 1.647538& $21.00\pm 0.37$& $3454\pm 50$& $2.40\pm 0.05$\\
\ldots          & \ldots             & \ldots   & \ldots  & \ldots         & \ldots      & \ldots        \\ 
\ldots          & 3222249533938897792& 82.045942& 1.947625& \ldots         & \ldots      & \ldots        \\
\ldots          & 3222249529644954880& 82.045953& 1.948293& \ldots         & \ldots      & \ldots        \\
\ldots          & 3222029008843059328& 82.143316& 1.448258& \ldots         & \ldots      & \ldots        \\
\ldots          & 3223738646345520000& 82.156917& 1.946033& \ldots         & \ldots      & \ldots        \\
\ldots          & 3223737615553348224& 82.160157& 1.906050& \ldots         & \ldots      & \ldots        \\
\noalign{\smallskip}
\hline
\noalign{\bigskip}
\end{tabular}
\begin{tabular}{cccccccccc}
\hline\hline
\noalign{\smallskip}
$\varpi$& $\mu_{\alpha*}$& $\mu_\delta$   & ruwe& astr. flag\tablefootmark{a}& $G$& 
  $G_\mathrm{BP}$& $G_\mathrm{RP}$& phot. flag\tablefootmark{a}& Prob.\tablefootmark{b}\\
(mas)   & (mas yr$^{-1}$)& (mas yr$^{-1}$)&     &  & (mag)& (mag)        & (mag)          &  &     \\
\hline
\noalign{\smallskip}
$2.712\pm 0.123$& $1.444\pm 0.124$& $-0.807\pm 0.095$& 1.01& 1& 17.670& 19.556& 16.383& 1& 0.997\\
$2.856\pm 0.060$& $1.480\pm 0.062$& $-0.087\pm 0.046$& 1.03& 1& 16.577& 18.101& 15.374& 1& 1.000\\
$2.853\pm 0.061$& $1.299\pm 0.065$& $-0.241\pm 0.048$& 1.02& 1& 16.663& 18.325& 15.468& 1& 1.000\\
$2.854\pm 0.019$& $1.228\pm 0.019$& $ 0.085\pm 0.014$& 1.04& 1& 14.105& 14.873& 13.230& 1& 1.000\\
$2.880\pm 0.045$& $1.410\pm 0.048$& $ 0.138\pm 0.035$& 1.02& 1& 16.012& 17.465& 14.822& 1& 1.000\\
\ldots          & \ldots          & \ldots   &  \ldots& \ldots& \ldots& \ldots& \ldots& \ldots& \ldots\\
$2.808\pm 0.053$& $1.207\pm 0.049$& $-0.098\pm 0.039$& 1.01& 1& 16.383& 18.047& 15.095& 1& 0.995\\
$3.141\pm 0.142$& $1.147\pm 0.130$& $ 0.027\pm 0.105$& 1.03& 1& 17.946& \ldots& \ldots& 1& 0.985\\
$2.928\pm 0.092$& $1.037\pm 0.122$& $-0.280\pm 0.096$& 0.95& 1& 17.154& 18.978& 15.888& 1& 0.995\\
$2.745\pm 0.281$& $1.629\pm 0.295$& $-0.268\pm 0.239$& 0.98& 1& 19.317& 21.196& 17.951& 0& 0.982\\
$2.941\pm 0.094$& $1.805\pm 0.081$& $-0.726\pm 0.070$& 1.01& 1& 17.223& 18.953& 15.953& 1& 0.982\\
\noalign{\smallskip}
\hline
\end{tabular}
\tablefoot{
\tablefoottext{a}{The astrometric and photometric flags are equal to 1, 0, or
$-1$ if the data are good, less reliable, or missing, respectively.}
\tablefoottext{b}{Membership probability from the RV and astrometric
analysis.}
}
\end{table*}

\end{appendix}

\end{document}